\documentclass[aps,amsmath,amssymb,prc,twocolumn,showpacs]{revtex4-2}

\usepackage{hyperref}
\usepackage{graphicx}
\usepackage{lineno}
\usepackage{dcolumn}
\usepackage{bm}
\usepackage{epstopdf}
\usepackage{amsmath,amssymb}

\newcommand{\snn}{\sqrt{s_{\mathrm {NN}}} = 200\mathrm{~GeV}}

\newcommand{\pt}{p_\mathrm{T}}
\newcommand{\kt}{KE_\mathrm{T}}
\newcommand{\vtwo}{v_\mathrm{2}}

\begin{document}

\title{Elliptic flow of strange and multi-strange hadrons in isobar collisions at $\snn$ at RHIC}
\author{The STAR Collaboration}


\begin{abstract}
We report a systematic measurement of elliptic flow ($v_{2}$) of $K_{s}^{0}$, $\Lambda$, $\overline{\Lambda}$, $\phi$, $\Xi^{-}$, $\overline{\Xi}^{+}$, and $\Omega^{-}$+$\overline{\Omega}^{+}$ at mid-rapidity ($|y| < 1.0$) for isobar, $^{96}_{44}$Ru+$^{96}_{44}$Ru and $^{96}_{40}$Zr+$^{96}_{40}$Zr, collisions at $\sqrt{s_{\mathrm {NN}}}$ = 200 GeV. The transverse momentum ($\pt$) dependence of $\vtwo$ is studied for various centrality classes. The number of constituent quark scaling of (multi-)strange hadrons is found to hold approximately within 20\%, suggesting the development of partonic collectivity in isobar collisions similar to that observed in Au+Au collisions at $\snn$. The average $\vtwo$ ratio shows $\sim$2\% deviation from unity in central and mid-central collisions for strange hadrons, indicating a difference in nuclear structure and deformation between the isobars. The $v_{2}$ in isobaric collisions is compared to Cu+Cu, Au+Au, and U+U collisions at similar collision energies. We observe an increase in $\vtwo$($\pt$) with increasing system size. The difference in $\vtwo$ between the isobar and other collision systems increases with $\pt$. The results are compared with a multi-phase transport model calculations with a deformed density profile to provide further insight into the nuclear structure of these isobars.
\end{abstract}


\maketitle

\section{INTRODUCTION}
Theoretical predictions from quantum chromodynamics (QCD) indicate that at sufficiently high temperatures and energy densities, there is a transition from hadron gas to a deconfined state of quarks and gluons known as Quark-Gluon Plasma (QGP)~\cite{QCD-1, QCD-2, lQCD}. Various observables like collective flow, particle spectra, nuclear modification factor, and strangeness production have been used to probe the QGP produced in high-energy nuclear collisions~\cite{ExQGP1,ExQGP2,ExQGP4,ExHIC1,ExQGP5,ExQGP6,LHCrev1,rev2}. Specifically, collective flow has been extensively studied to comprehend the properties of the QGP matter~\cite{flow4,flowmm2,STARsys,flowm1,flowm3}. Collective behavior due to the strength of interactions within the collision medium turn initial spatial anisotropies into final-state momentum-space anisotropies which can be measured in experiments~\cite{flowm5}. The spatial anisotropies may be caused by asymmetry of the initial overlap region, fluctuations in the position of the participant nucleons~\cite{flucPHOBOS}, and structure and deformation of the colliding nuclei~\cite{DFT-1,structure,structure2}. The collective flow can be measured via the Fourier expansion of the azimuthal angle ($\phi$) distribution of the produced particles with respect to the $n^{th}$-order flow symmetry plane, $\Psi_{n}$, as $v_{n} =\langle \cos n(\phi - \Psi_{n})\rangle$~\cite{flow1}. The second-order flow coefficient, $\vtwo$, known as elliptic flow, provides valuable information about pressure gradients in a hydrodynamic framework, the extent of thermalization, the equation of state of the matter formed and the effective degrees of freedom in the matter produced by the collisions~\cite{hydroOllit,flowm4,revFlow}. The self-quenching nature of $\vtwo$ makes it particularly sensitive to early stage collision dynamics~\cite{flowm4,flow2,flow4}. In particular, the strange and multi-strange hadrons have small hadronic interaction cross-section and are expected to freeze out earlier than the light hadrons, at a temperature nearer to the transition from the QGP~\cite{phi1,phi2,phi3,phi4}. Therefore, $\vtwo$ measurements of these hadrons are less affected by late-stage hadronic interactions and provide more information on the early-stage collision dynamics~\cite{flow2,AuCent}.

A run of isobaric collisions, $^{96}_{44}$Ru+$^{96}_{44}$Ru and $^{96}_{40}$Zr+$^{96}_{40}$Zr at center-of-mass energy $\snn$ was successfully carried out at the Relativistic Heavy-Ion Collider (RHIC) in 2018 with the chief goal of measuring the charge separation along the magnetic field, known as the Chiral Magnetic Effect (CME)~\cite{STARisobarData,expProp,ReviewCME}. The difference in the proton number between the two isobar nuclei results in a higher magnetic field, which was anticipated to show a larger CME-related signals in Ru+Ru collisions compared to Zr+Zr collisions. Apart from the CME exploration study, the results from the isobaric collisions have shown an interesting observation of difference in the elliptic and triangular flow of charged hadrons between the two isobars despite the same mass number. This result suggests a different nuclear density and deformation of the nuclei under consideration~\cite{STARisobarData}. These measurements have been used to decipher the nuclear structure of these isobars in various model calculations. A wealth of studies using transport and hydrodynamic models with different Woods-Saxon (WS) parameterization for isobars have demonstrated the effects of nuclear deformation and neutron thickness on the particle production, collective flow ($v_{2}$ and $v_{3}$), and $\vtwo$-$\langle p_\mathrm{T} \rangle$ correlations~\cite{myampt,amptLN,quadAndOctoFlow,v2AndShape1,v2Isobar2,meanPt1,neutronSkin,neutronSkin2,trentoNeutronskin,trentoUrQMD,DFT-1,DFT-2,Trajectum}. A systematic study of elliptic flow of (multi-)strange hadrons in these small and deformed nuclei, Ru+Ru and Zr+Zr will give more information to help understand the nuclear structure.

Recent results show QGP-like signatures in small collision systems, which warrants investigating the dynamics and evolution of the medium produced in these collisions~\cite{QGPsmall1,STARdAu,STARsmall,PHENIXsys,QGPsmall2,QGPsmall3,QGPsmall4,QGPsmall5}. One such signature, strangeness enhancement in small collision systems, suggests a common mechanism for strange particle production across varying collision system sizes, which may be linked to the formation of a QGP medium~\cite{NatureALICE}. Additionally, the dependence of the anisotropic flow of charged and light hadrons on system size has been investigated for a wide range of nuclei involved in heavy-ion collisions such as U+U, Pb+Pb, Au+Au, Xe+Xe, Cu+Au, and Cu+Cu, as well as in smaller collision systems like p+p, p+Au, d+Au, $\mathrm{^{3}He}$+Au, and p+Pb~\cite{STARsys,flowe3,STARUU,PHENIXsys,ALICEsys}. These studies reveal a system-size hierarchy of $\vtwo$ resulting from the medium response to the eccentricity of the initial overlap region of the colliding nuclei. Therefore, comparing the elliptic flow of strange and multi-strange hadrons in small and deformed nuclei, such as Ru+Ru and Zr+Zr, across different colliding systems can further aid our understanding of how system size affects the produced medium.

We report a systematic measurement of elliptic flow of strange and multi-strange hadrons as a function of $\pt$ and collision centrality in isobar, Ru+Ru, and Zr+Zr, collisions at $\snn$ from the STAR experiment at RHIC. The paper is organized as follows. We discuss the STAR detector setup, event and track selection, particle identification and analysis technique in Sec.~\ref{experiment}. The systematic uncertainties associated with the measurements are discussed in Sec.~\ref{sys_error}. The results of $\pt$ and centrality dependence of $\vtwo$ measurements for $K_{s}^{0}$, $\Lambda$, $\overline{\Lambda}$, $\phi$, $\Xi^{-}$, $\overline{\Xi}^{+}$, and $\Omega^{-}$+$\overline{\Omega}^{+}$ are presented in Sec.~\ref{result}. The number of constituent quark scaling, system size dependence, and comparison to transport model calculations is discussed in Secs.~\ref{ptdep}-\ref{model}. Finally, a summary of the results reported in this paper is given in Sec.~\ref{summary}.

\section{EXPERIMENTAL SETUP AND ANALYSIS TECHNIQUE}
\label{experiment}
\subsection{STAR Detector System}
\label{star}
The Solenoidal Tracker at RHIC~\cite{RHIC1,STARd1}, abbreviated as STAR, is one of the experiments running at RHIC. Various light and heavy-ions like O+O, Cu+Cu, Ru+Ru, Zr+Zr, Au+Au, and U+U are collided at RHIC at multiple center-of-mass energies scanning a wide range in the QCD phase diagram~\cite{starwhite,starbesreview,bes1,bes2,bes3,bes4,QCDphase1,QCDphase2}. The datasets used in this analysis are Ru+Ru and Zr+Zr collisions at $\sqrt{s_{NN}}$ = 200 GeV from the year 2018. The main sub-systems used in this measurement are the Time Projection Chamber (TPC) and the Time-Of-Flight (TOF)~\cite{TPC,TOF}. The Vertex Position Detector (VPD) is used as a trigger to record minimum bias collision events, which are defined by a coincidence between the East and West VPD~\cite{STARvpd1,STARtrigger1}. The TPC is the primary detector used in the STAR experiment for particle tracking and identification. It identifies particles based on average energy loss per unit length due to ionization ($\langle dE/dx \rangle$). The TOF detector is used to identify particles at higher momentum using the time of flight for particles to reach the detector, combined with momentum and path-length information from the TPC. The start time for the TOF detector is given by the VPD system. Both TPC and TOF detectors are used in this analysis to identify pions, kaons, and protons.

\subsection{Event Selection}
\label{evtrk}
Events included in this study have a reconstructed collision vertex (primary vertex) located within a $-35 < V_{z} < 25$ cm range from the center of the TPC detector along the beam direction. The asymmetry in the $V_{z}$ results from the online vertex selection procedure, stemming from the timing offset between VPD east and west. This criterion ensures a roughly uniform detector acceptance. Additionally, the events are required to have a primary vertex within 2 cm of the nominal collision point in the radial direction ($V_{r} = \sqrt{V_{x}^{2} + V_{y}^{2}}$) to exclude events resulting from the beam-pipe interaction. The VPDs also provide collision-vertex ($V_{z,VPD}$) to select collision events near the center of the detector. An additional selection on the difference between the $V_{z}$ positions determined by the TPC and VPD ($|V_{z} - V_{z,VPD}| <$ 5 cm) is applied to discard pile-up events.  

The centrality of an event is based on the multiplicity of charged particles having pseudorapidity $|\eta|<$ 0.5, distance of closest approach to the primary vertex (DCA) $<$ 3 cm, and at least ten ionization points ($N_{fits}^{hit}$) within the TPC. The charged particle multiplicity is corrected for a detection efficiency that depends on beam luminosity and the acceptance variation as a function of $V_{z}$. This charged particle multiplicity from the TPC is fitted with the multiplicity obtained from two-component nucleon-based Monte-Carlo Glauber simulation~\cite{glauber,centrality1,STARcent}, the details of which can be found in Ref.~\cite{STARisobarData}. The data set are divided into nine centrality classes, viz., 0-5\%, 5-10\%, 10-20\%, and so on upto 70-80\%. After the event and centrality selection, the total number of minimum bias (MB) events analyzed are $\sim$1.77 billion for Ru+Ru and $\sim$1.84 billion for Zr+Zr collisions at $\snn$.

\subsection{Track selection and Particle Identification}
\label{particle_identifcation}
\begin{figure}[!htbp]
\begin{tabular}{cc}
\includegraphics[width=0.22\textwidth]{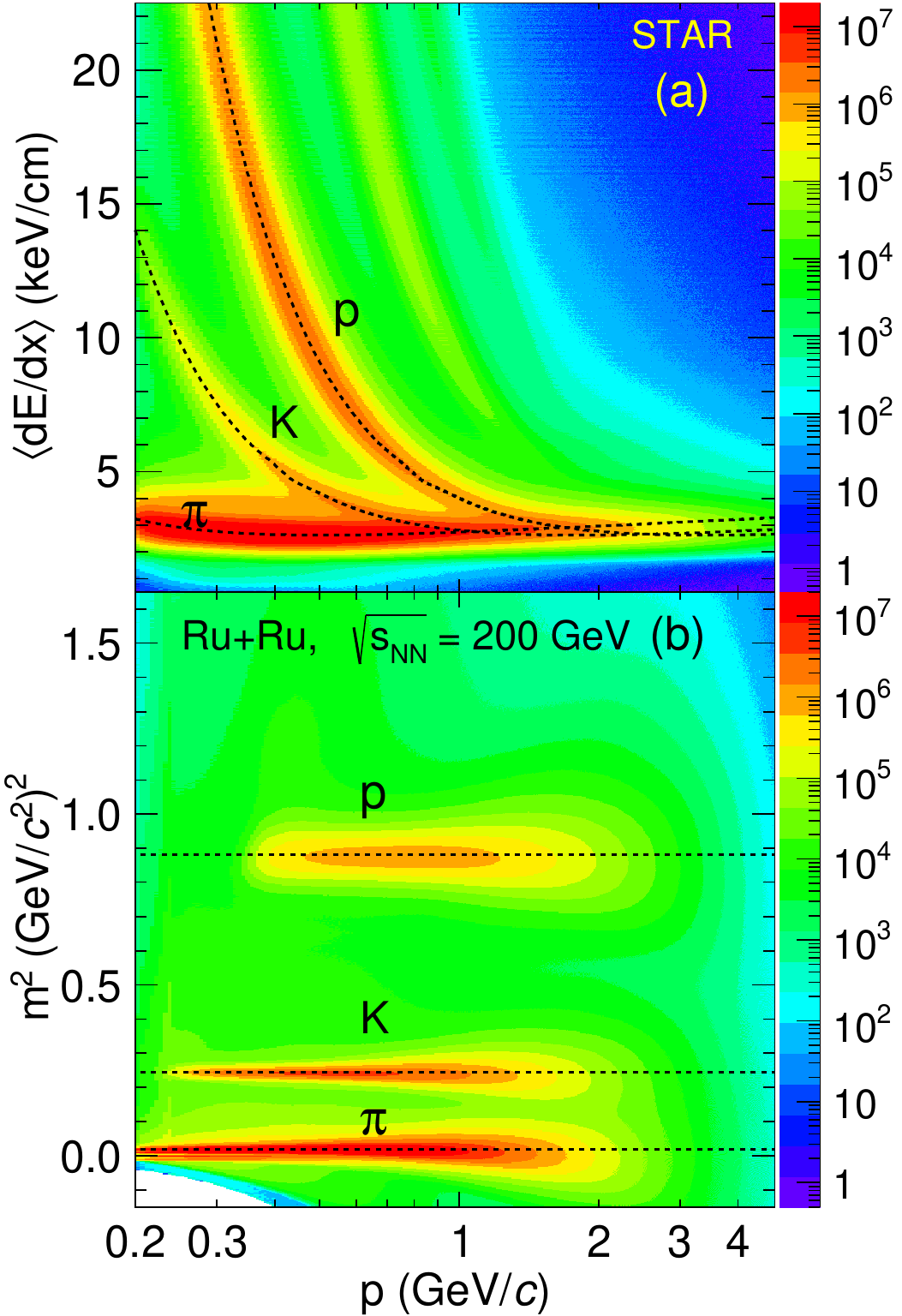} & \includegraphics[width=0.22\textwidth]{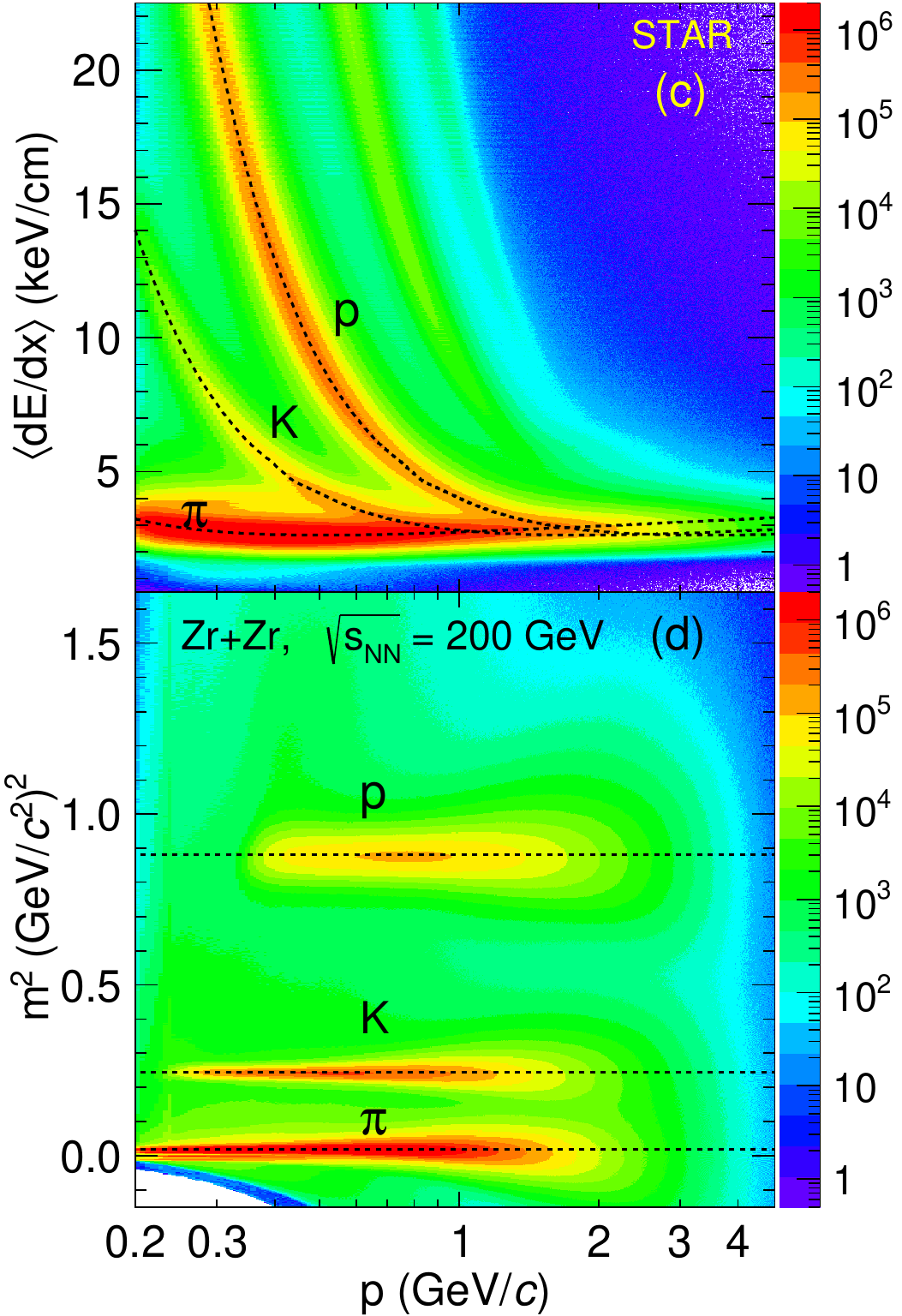}
\end{tabular}
\caption{(a) The $\left\langle dE/dx \right\rangle$ distribution of particles identified using the TPC as a function of momentum (p) within $|\eta| <$ 1.0 for (a) Ru+Ru and (c) Zr+Zr collisions at $\sqrt{s_{NN}}$ = 200 GeV. The dashed lines represent the theoretical $\left\langle dE/dx \right\rangle$ values for the corresponding particle. Bottom panels: mass square as a function of p using TOF information in (b) Ru+Ru and (d) Zr+Zr collisions at $\sqrt{s_{NN}}$ = 200 GeV. The dashed lines represent the mass square values from PDG for the corresponding particle.}
\label{fig:dedxm2}
\end{figure}

We reconstruct the short-lived particles, $K^{0}_{s}$, $\phi$, $\Lambda$, $\Xi$, and $\Omega$ via their hadronic decay channels as listed in Tab.~\ref{tab:liststrange}~\cite{pdg1}. These particles are reconstructed using the decay daughter tracks of $\pi^{\pm}$, $K^{\pm}$, and $p(\overline{p})$ within $|\eta| <$ 1.0. Only tracks with $N_{fit}^{hits}$ $\geq$ 15 are selected to ensure adequate momentum resolution. Additionally, a minimum $\pt$ of the tracks is chosen to be 0.2 GeV/$\it{c}$ to avoid short tracks traversing through the TPC. A DCA cut of less than 3 cm is applied on the kaon tracks to reduce the contamination of secondary tracks from weak decays in the reconstruction of $\phi$-mesons.
\begin{table}[!htbp]
    \centering
    \begin{tabular}{c|c|c}
    \hline
    \textbf{Particle} & \textbf{Decay channel} & \textbf{Branching ratio (\%)} \\ \hline
        $K^{0}_{s}$  & $\pi^{+}$ + $\pi^{-}$ & 69.2 \\
        $\phi$  & $K^{+}$ + $K^{-}$ & 49.2\\
        $\Lambda$($\overline{\Lambda})$  &  $p$ + $\pi^{-}$ ($\overline{p}$ + $\pi^{+})$ & 63.9 \\
        $\Xi^{-}$($\overline{\Xi}^{+})$  & $\Lambda$ + $\pi^{-}$ ($\overline{\Lambda}$ + $\pi^{+})$ & 99.887 \\
        $\Omega^{-}$($\overline{\Omega}^{+})$ & $\Lambda$ + $K^{-}$ ($\overline{\Lambda}$ + $K^{+}$) & 67.8 \\
        \hline
    \end{tabular}
    \caption{Decay channels and respective branching ratios of strange and multi-strange hadrons used in the analysis.}
    \label{tab:liststrange}
\end{table}
Their decay daughters, $\pi^{\pm}$, $K^{\pm}$, and $p(\overline{p})$ are identified by combining the normalized $\left\langle dE/dx \right\rangle$ from the TPC detector and the velocity ($\beta = L/ct$) from the TOF detector. The normalized $\left\langle dE/dx \right\rangle$ is defined as
\begin{equation}
\label{eq:nsigma}
  n\sigma_{\pi/K/p} = \frac{1}{R}\ln\left(\frac{\left\langle dE/dx\right\rangle_{measured}}{\left\langle dE/dx\right\rangle_{\pi/K/p}^{th}}\right),
\end{equation}
where $\left\langle dE/dx \right\rangle_{\pi/K/p}^{th}$ is the expected $\left\langle dE/dx \right\rangle$ calculated using the Bichsel function for the respective particle and $R$ is the experimental $\left\langle dE/dx \right\rangle$ resolution~\cite{bichsel}. Figure~\ref{fig:dedxm2} (a) and (c) show the $\left\langle dE/dx \right\rangle$ of charged particles as a function of momentum in Ru+Ru and Zr+Zr collisions at $\snn$. The dashed lines represent theoretical values given by parametrization of the Bichsel function. At higher momentum, the $\left\langle dE/dx \right\rangle$ values of different particles overlap. Therefore, the time-of-flight information of tracks from the TOF detector is used to identify particles at higher momentum ranges. Using the velocity information from TOF and momentum information from the TPC, the square of the mass ($m^{2}$) of the corresponding particle is calculated using the relation $m^{2} = p^{2}\left(1/\beta^{2} -1 \right)$. Figure~\ref{fig:dedxm2} (b) and (d) show the $m^{2}$ as a function of momentum. The dashed lines are the $m^{2}$ values for pions, kaons, and protons from the Particle Data Group~\cite{pdg1}.

The strongly decaying $\phi$-meson ($s\overline{s}$) is reconstructed using the invariant mass technique via its hadronic decay channel. The determination of uncorrelated combinatorial background for $\phi$-mesons is carried out using the mixed-event technique. Events with similar properties are grouped for mixing, reproducing the shape of background distribution well. The events are mixed based on the categories of nine bins of centrality (from 0-5\%, 5-10\%, 10-20\% up to 70-80\%), ten equal size bins of z-vertex ($V_{z}$), and six bins of event plane angle ($\psi_{2}$) in the range between 0 to $\pi$, resulting in a total of 540 event classes. In this method, $K^{+}$ from one event is combined with oppositely charged $K^{-}$ from five other events, which removes the correlation. The uncorrelated background is then scaled using the side-band method and subtracted from the invariant mass distribution of the same event. The remaining signal after background subtraction is fitted with a Breit-Wigner function plus a second-order polynomial to account for the residual background~\cite{phimeson2}.

The weakly decaying neutral strange particles $K_{s}^{0}$ and $\Lambda(\overline{\Lambda})$ are also reconstructed using the invariant mass technique. The neutral particles decay into two oppositely charged hadrons (forming a V-shaped decay topology) away from the primary vertex. The secondary vertex corresponding to the point of decay is reconstructed using the $V^{0}$ topology of the decay daughters. To select the daughter pions and protons of $K_{s}^{0}$ and $\Lambda$, a criterion of $|n\sigma_{\pi,p}| <$ 3.0 is applied. Additionally, when the TOF track information is available, a selection criterion on $m^{2}$ is utilized. The combinatorial background is constructed using the rotational background method where one of the daughter tracks is rotated by $180^{\circ}$ in the transverse plane, thereby breaking the correlation between the daughter tracks~\cite{thesis_paul}. 

The multi-strange particles $\Xi^{-}(\overline{\Xi}^{+})$ and $\Omega^{-}(\overline{\Omega}^{+})$ have a two-step decay. First, they decay into a charged and a neutral $V^{0}$ particle ($\Lambda$/$\overline{\Lambda}$), which then decays into two charged particles. These multi-strange baryons are reconstructed by calculating the decay kinematics of the three charged decay daughters. The first reconstruction step is to find the decay vertex of neutral $V^{0}$ hadron using decay kinematics. After finding a suitable $V^{0}$ candidate, the next step is to search for a matching charged pion or kaon from $\Xi$ or $\Omega$ decay using the topological cuts. For $\Xi$ and $\Omega$ particles, the momentum vector of the charged daughter (pion or kaon) is rotated to estimate the background distribution~\cite{idflowe1}. 

\begin{figure*}[!htbp]
\centering
\includegraphics[width=0.24\textwidth]{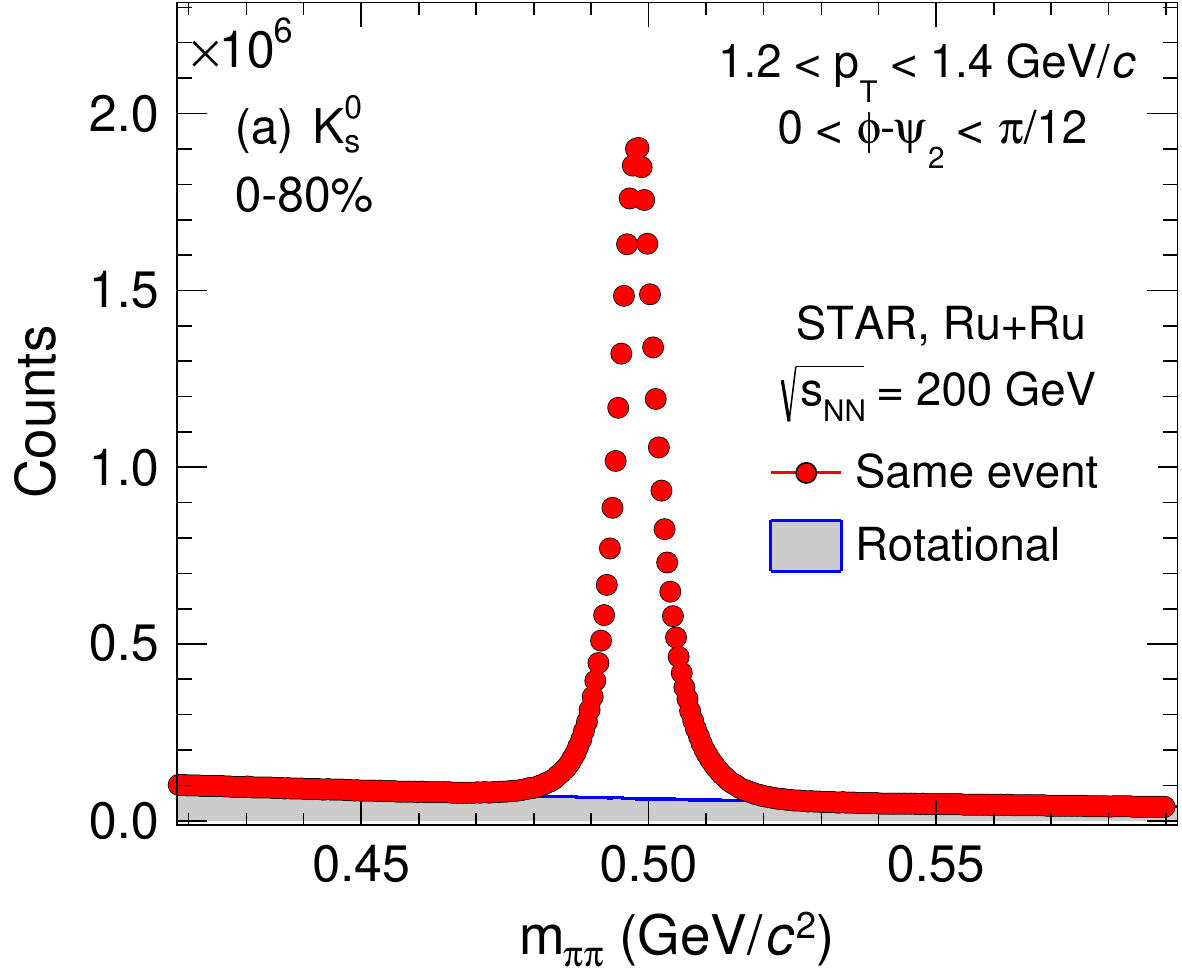} 
\includegraphics[width=0.24\textwidth]{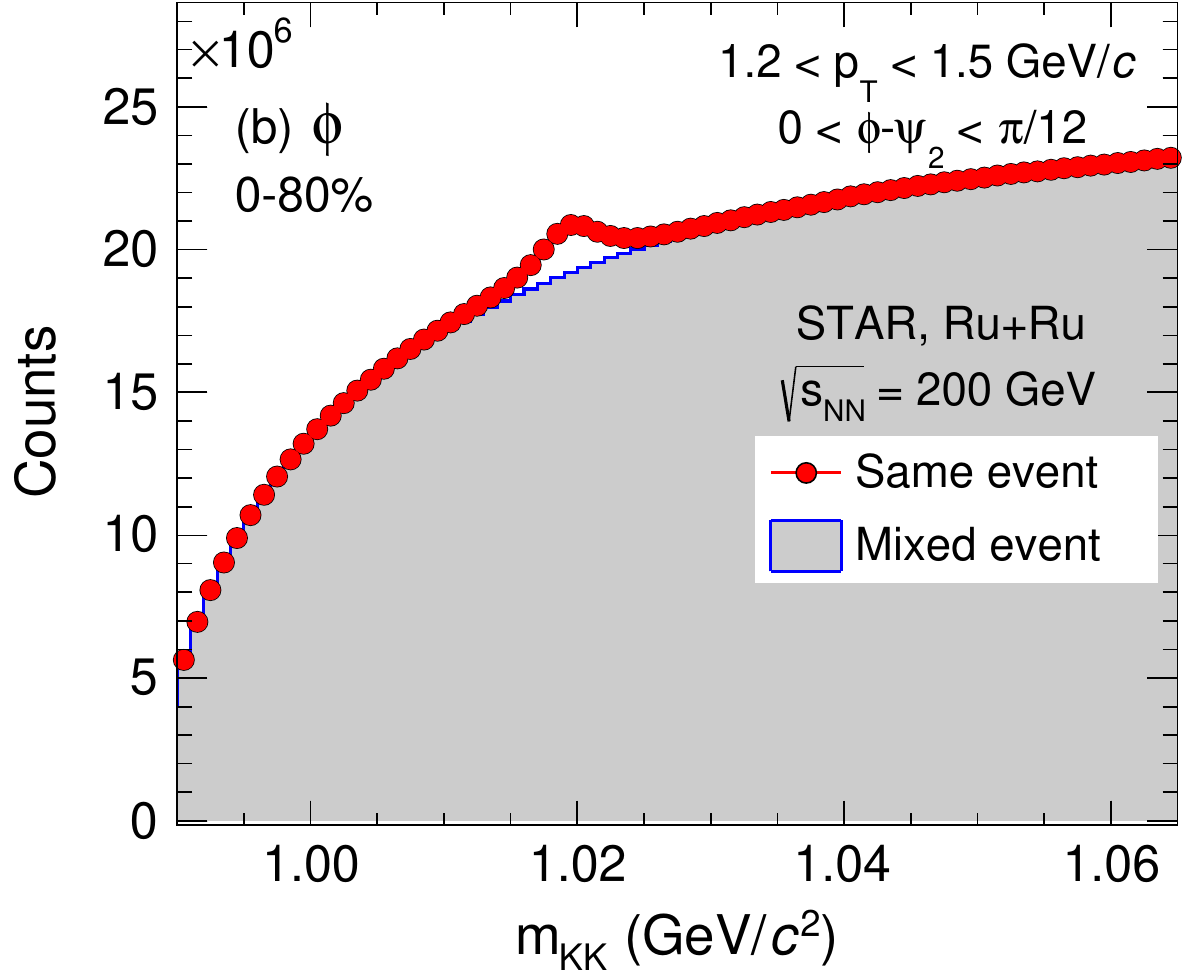}
\includegraphics[width=0.24\textwidth]{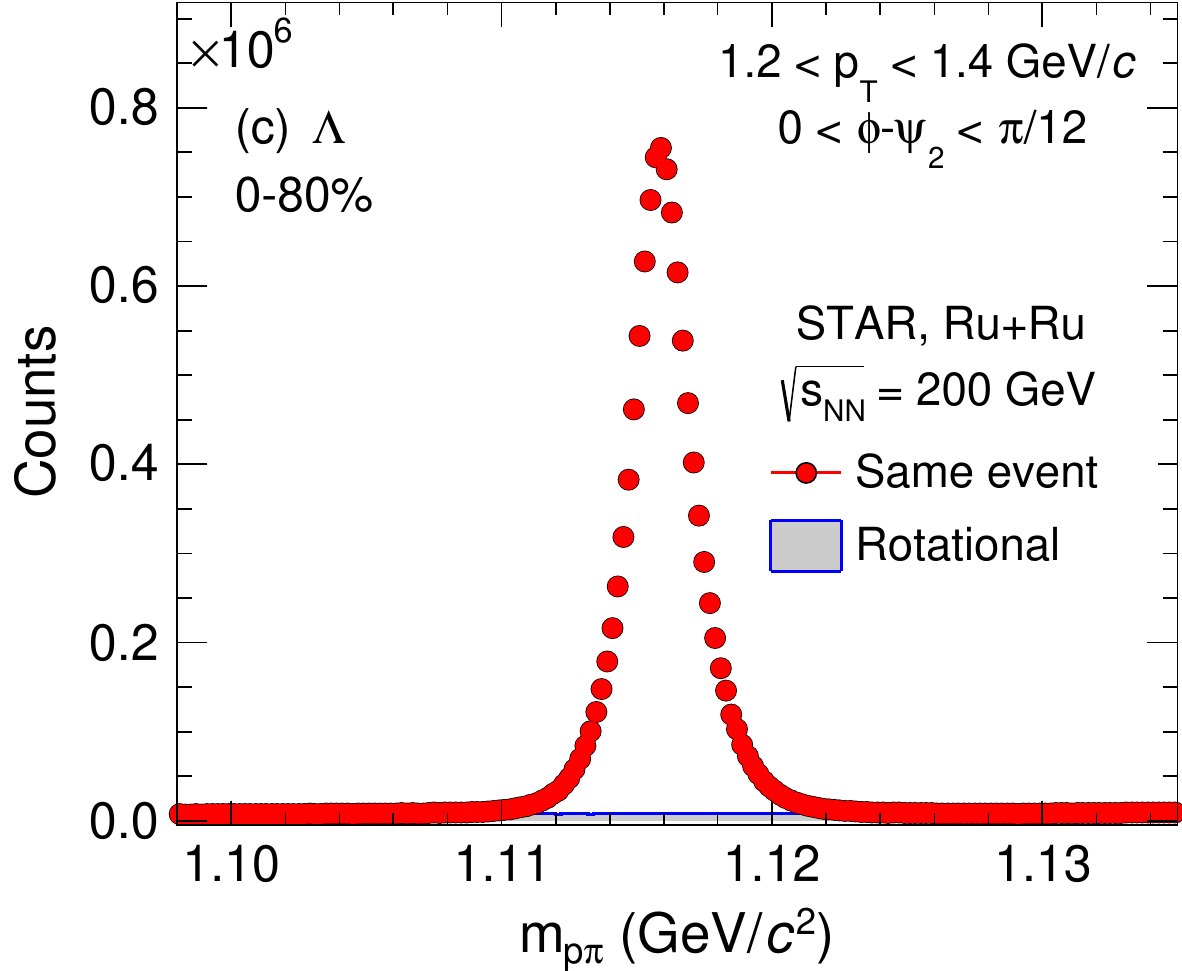}
\includegraphics[width=0.24\textwidth]{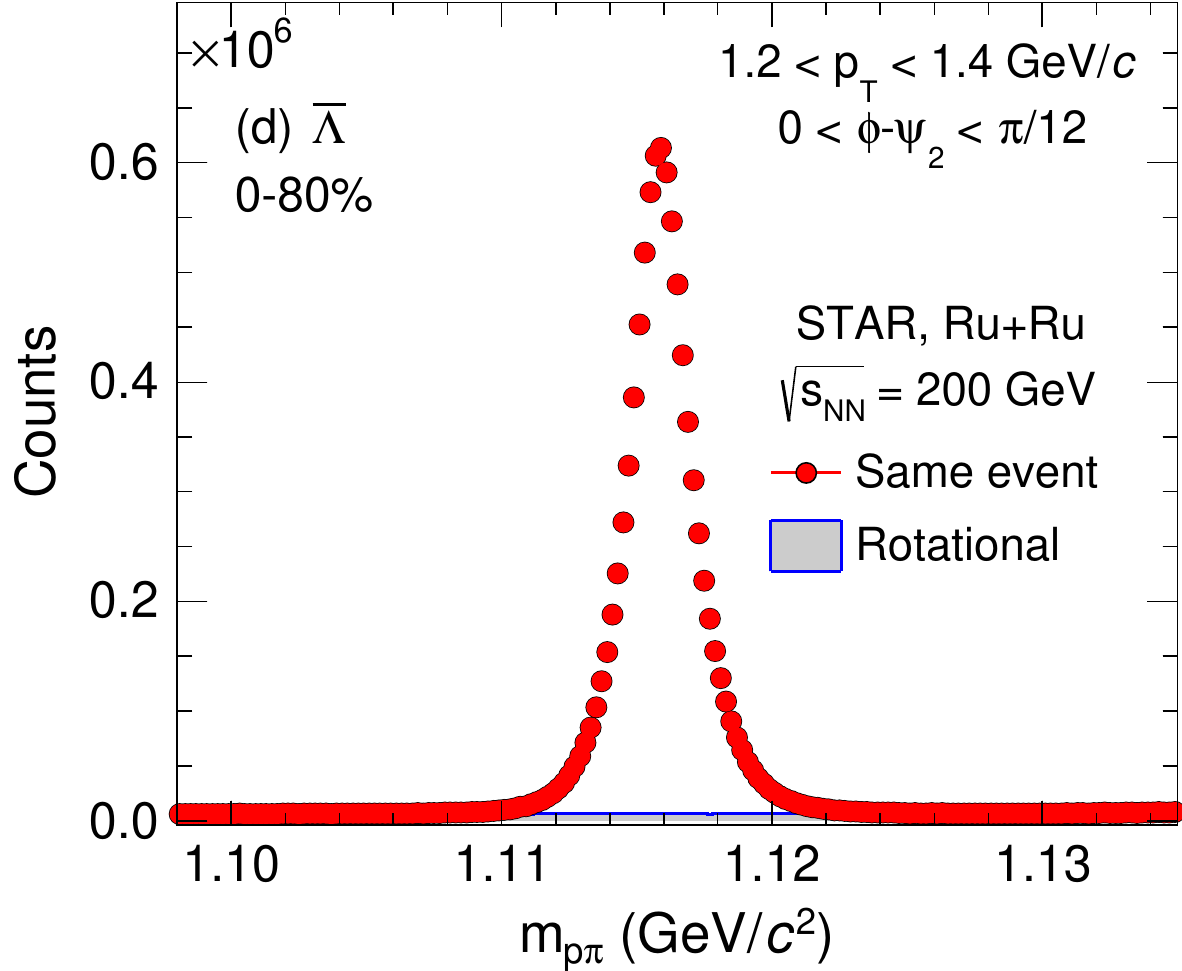} 
\includegraphics[width=0.24\textwidth]{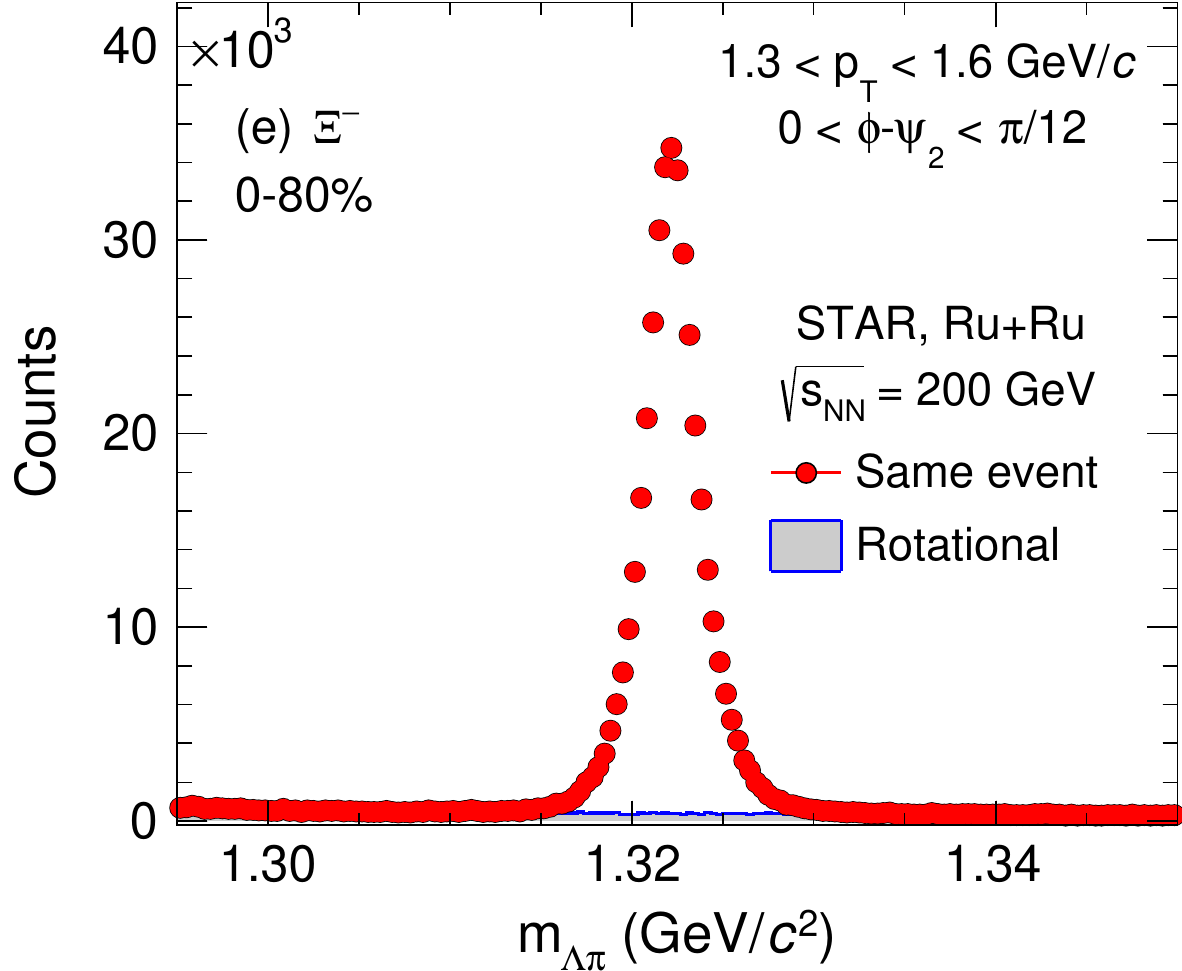}
\includegraphics[width=0.24\textwidth]{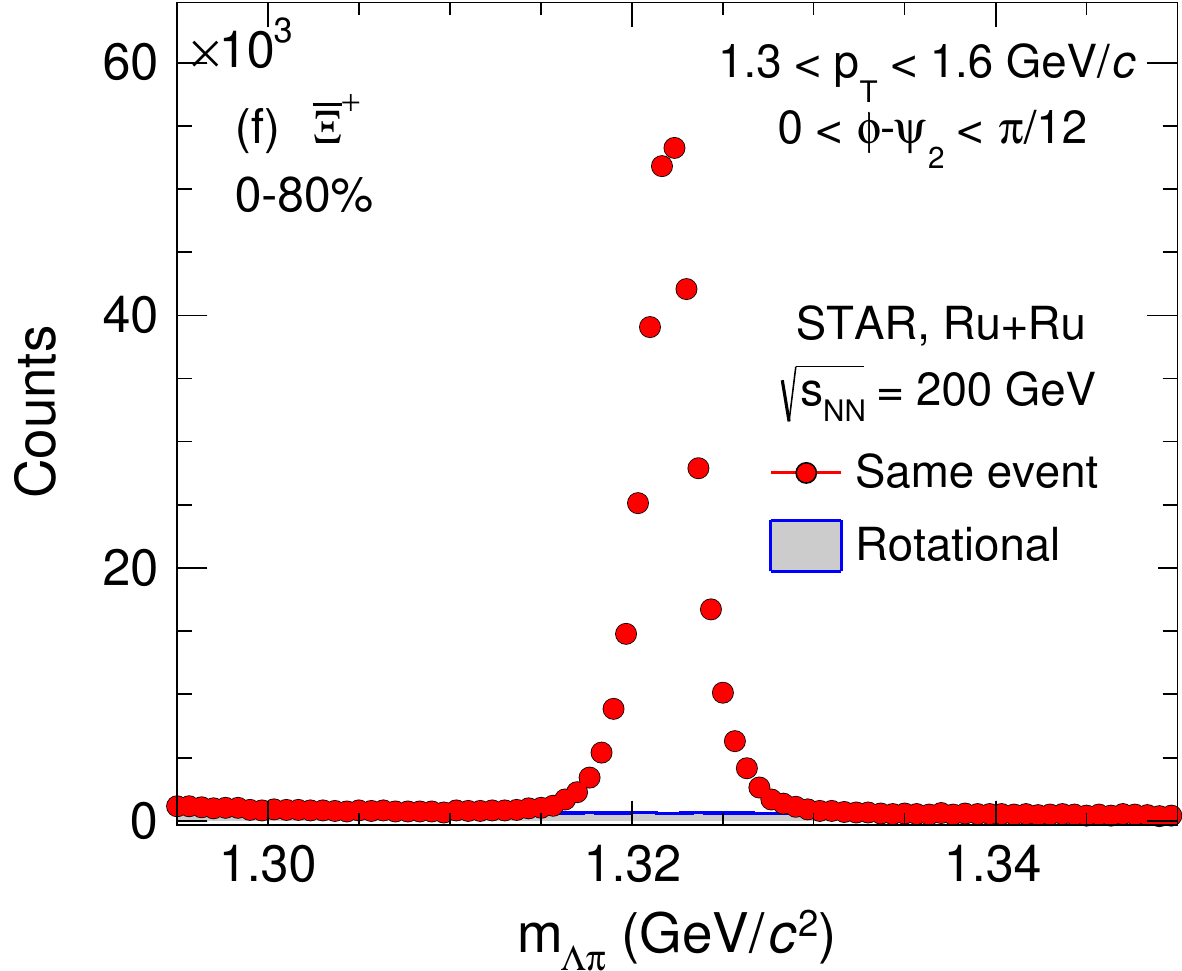}
\includegraphics[width=0.24\textwidth]{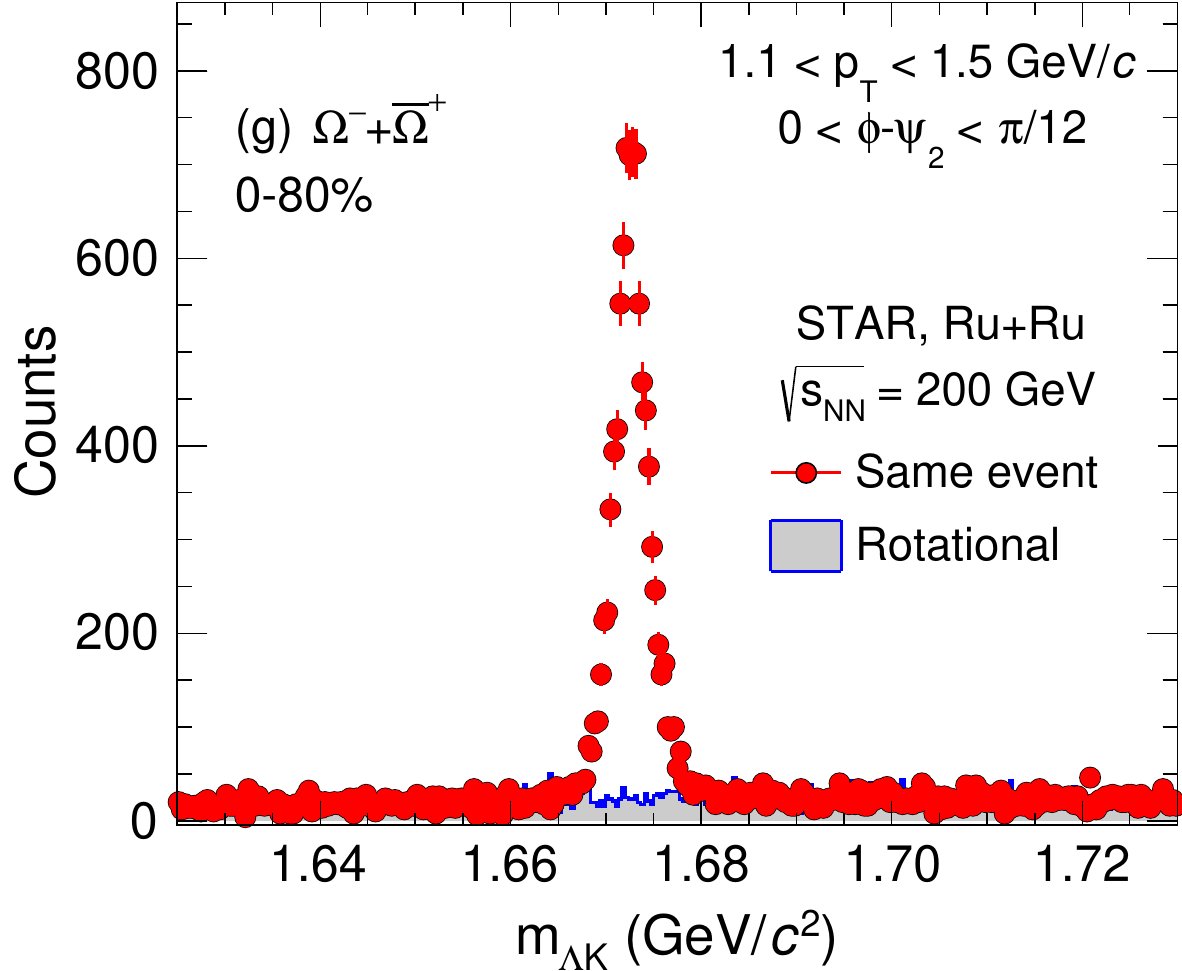}
\caption{Invariant mass distributions for (a) $K^{0}_{s}$, (b) $\phi$, (c) $\Lambda$, (d) $\overline{\Lambda}$, (e) $\Xi^{-}$, (f) $\overline{\Xi}^{+}$, and (g) $\Omega^{-}$+$\overline{\Omega}^{+}$ in minimum bias Ru+Ru collisions at $\sqrt{s_{NN}}$ = 200 GeV. The gray bands represent the combinatorial background. The rotational background technique is used for $K^{0}_{s}$, $\Lambda$, $\Xi$, and $\Omega$, while the mixed event technique is used for $\phi$ mesons. Error bars represent the statistical uncertainties.}
\label{fig:invmass1}
\end{figure*}

\begin{figure*}[!htbp]
\centering
\includegraphics[width=0.24\textwidth]{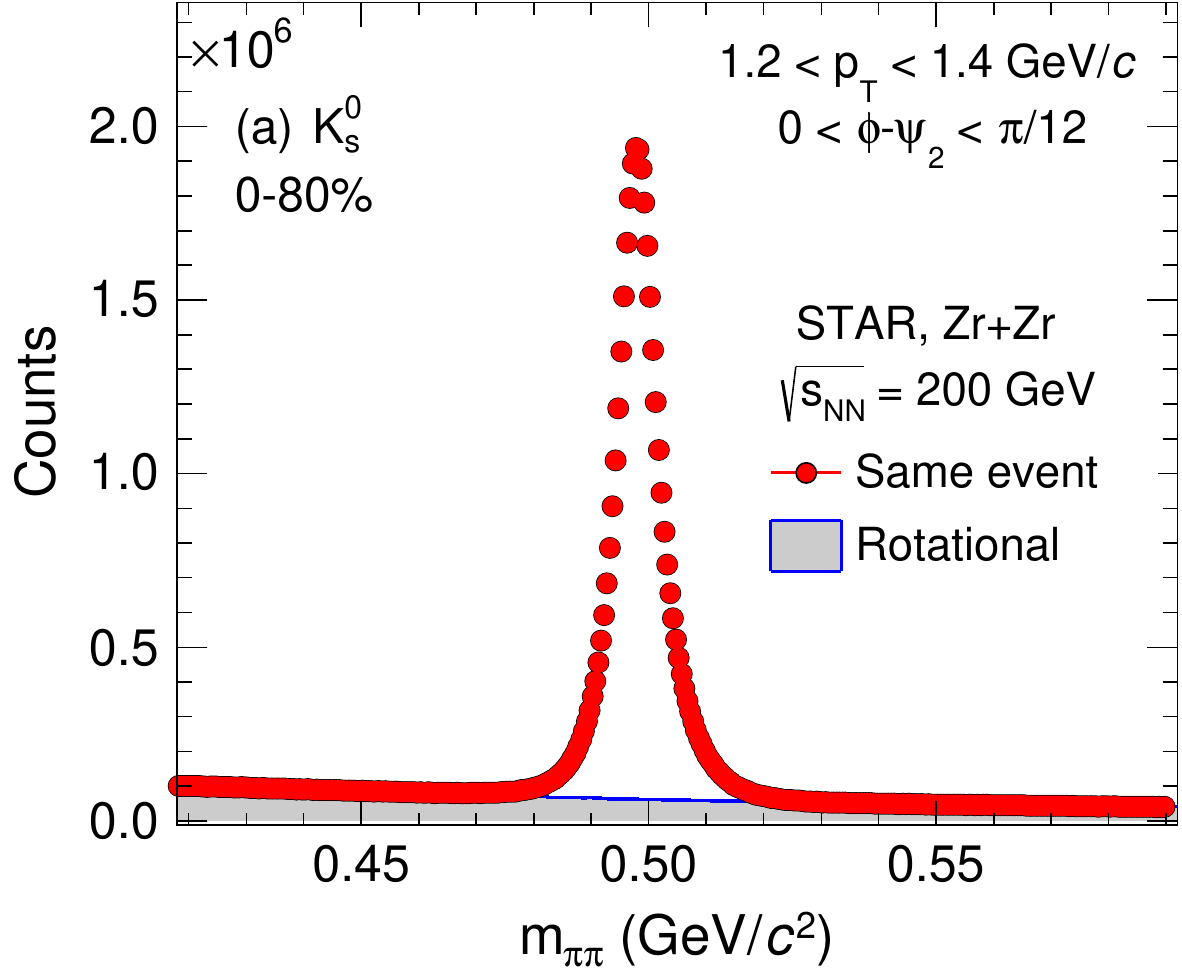} 
\includegraphics[width=0.24\textwidth]{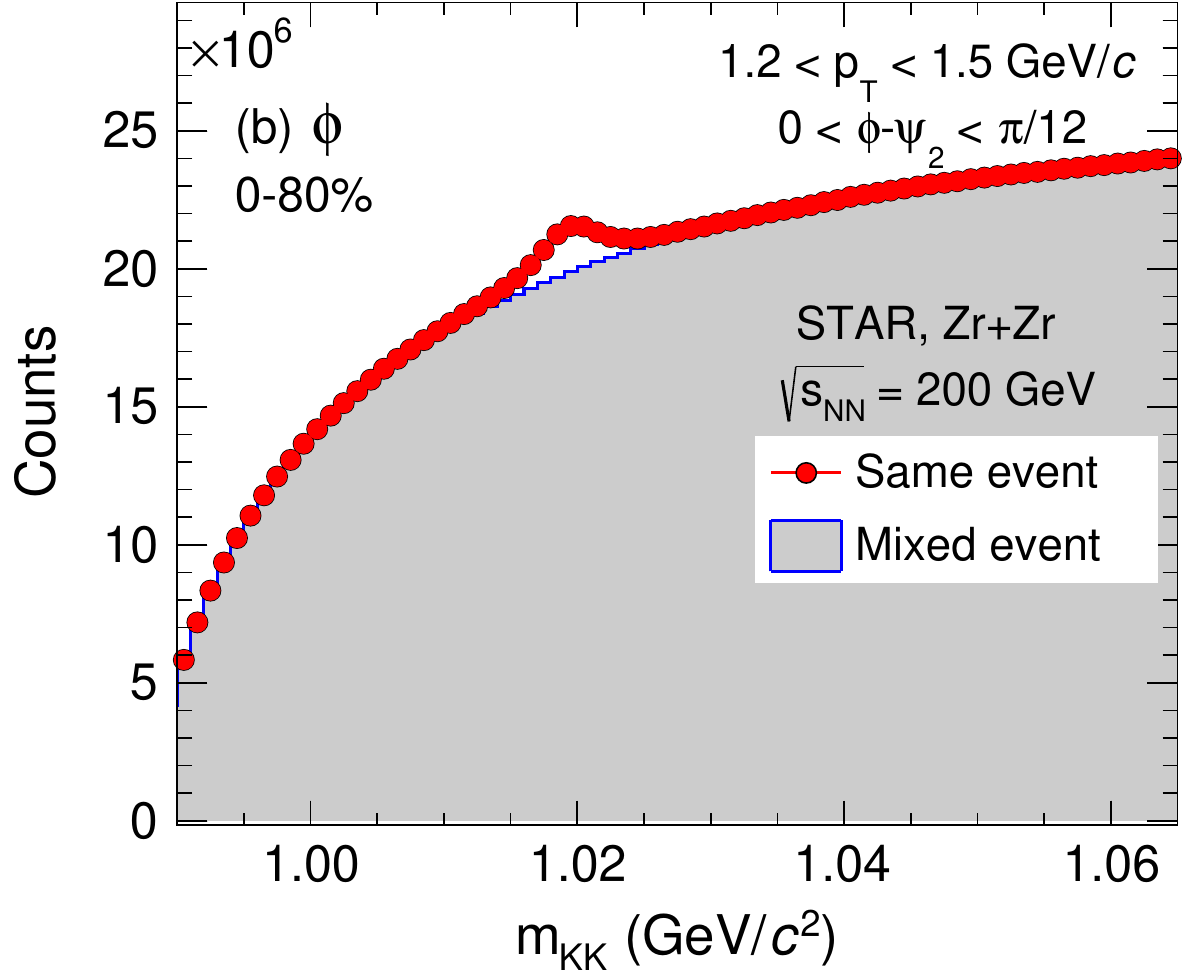}
\includegraphics[width=0.24\textwidth]{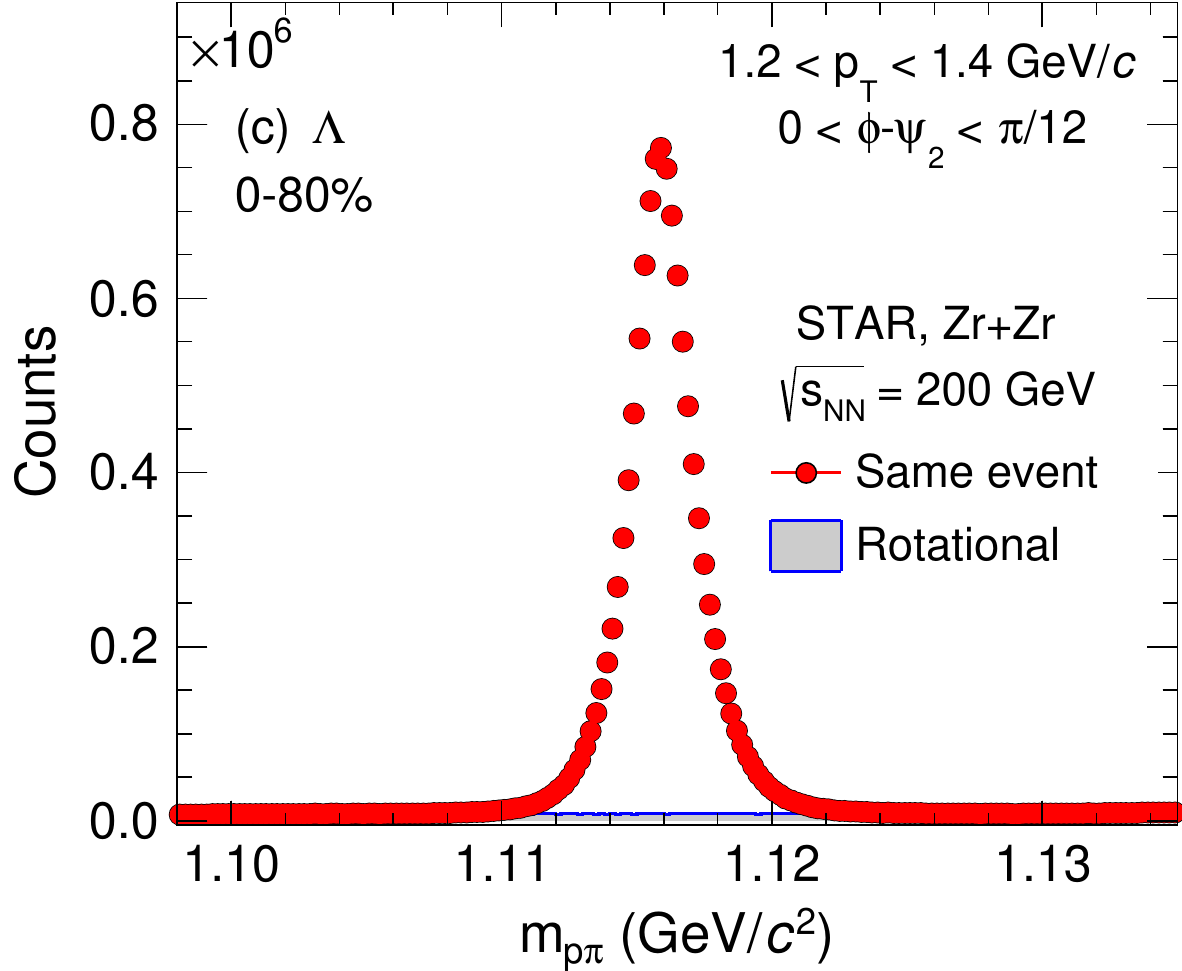}
\includegraphics[width=0.24\textwidth]{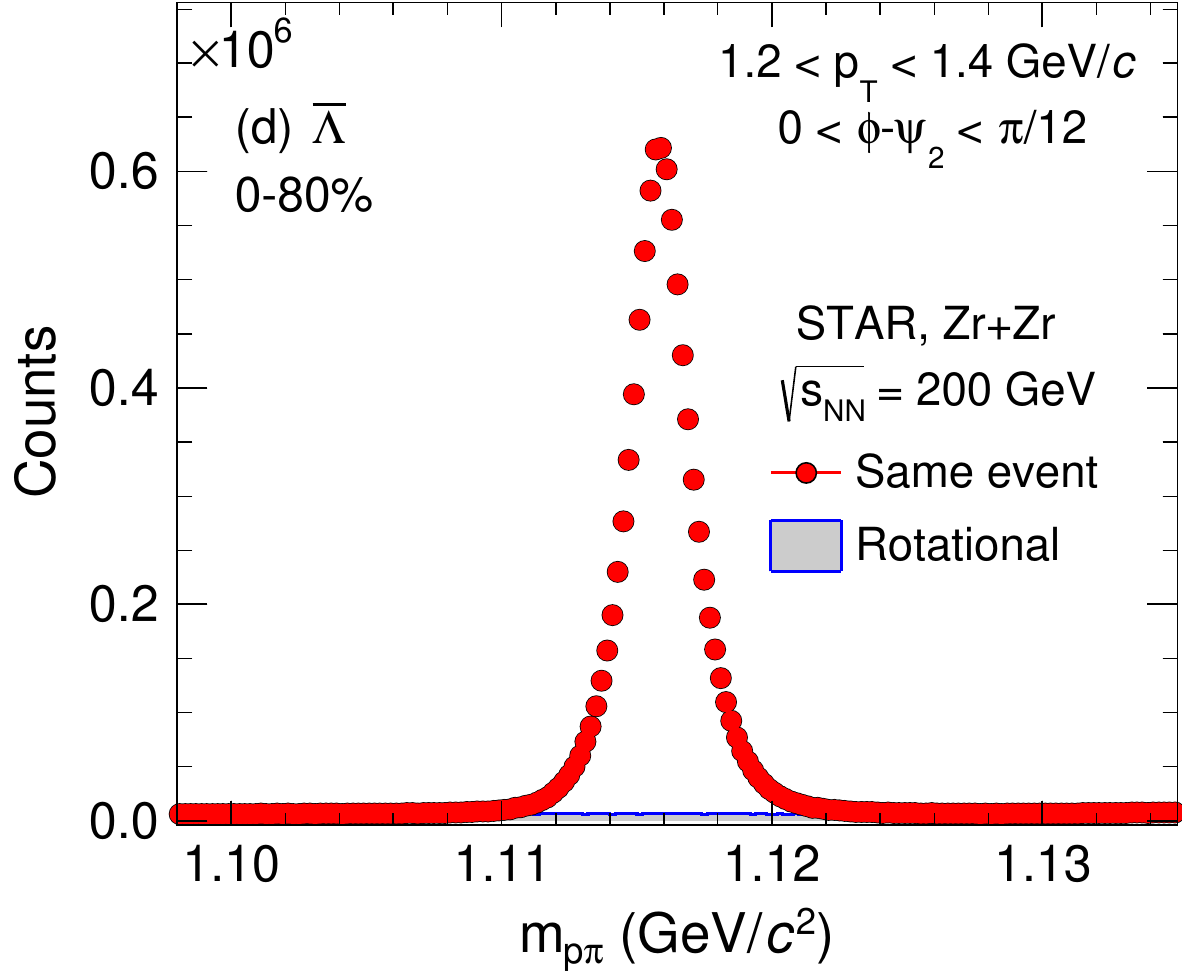} 
\includegraphics[width=0.24\textwidth]{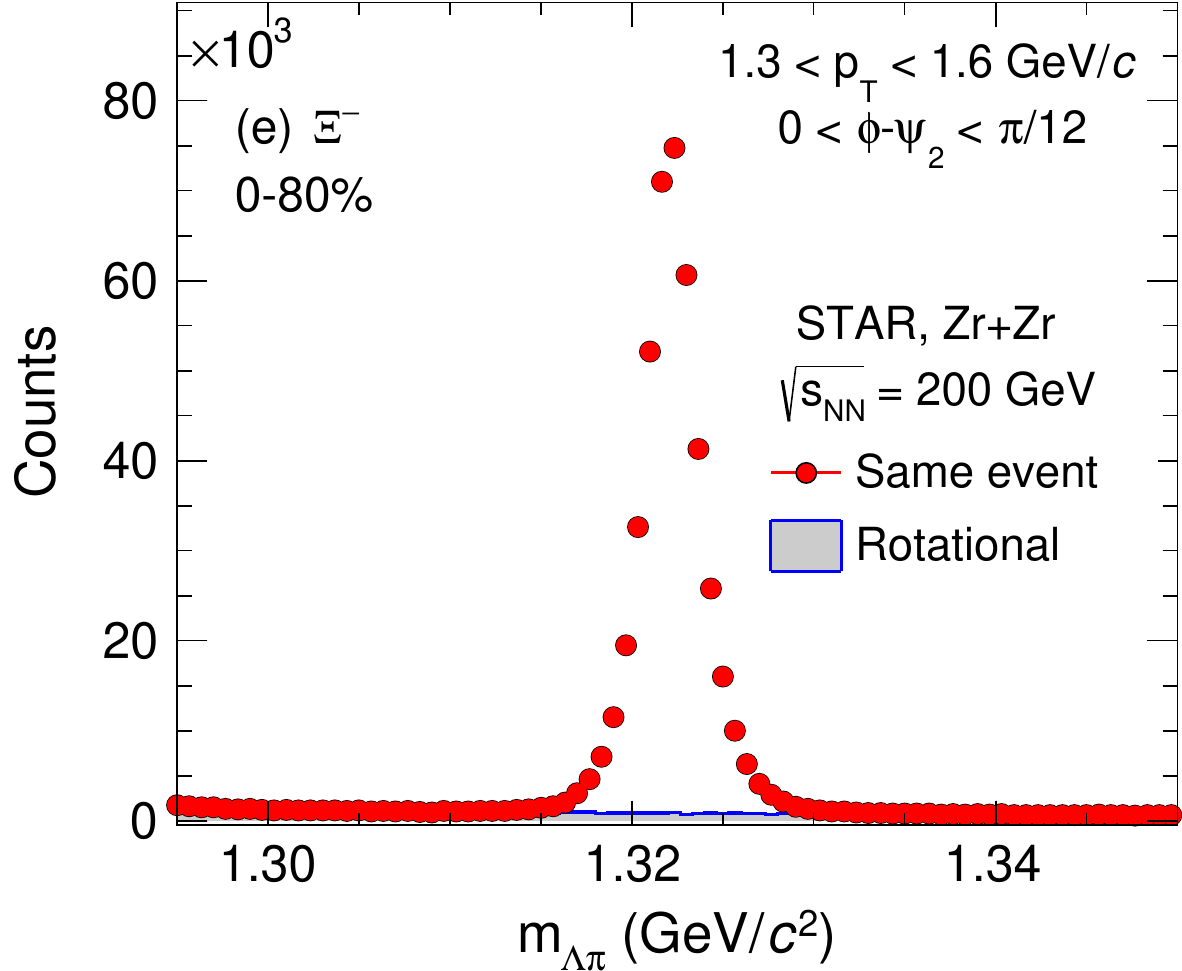}
\includegraphics[width=0.24\textwidth]{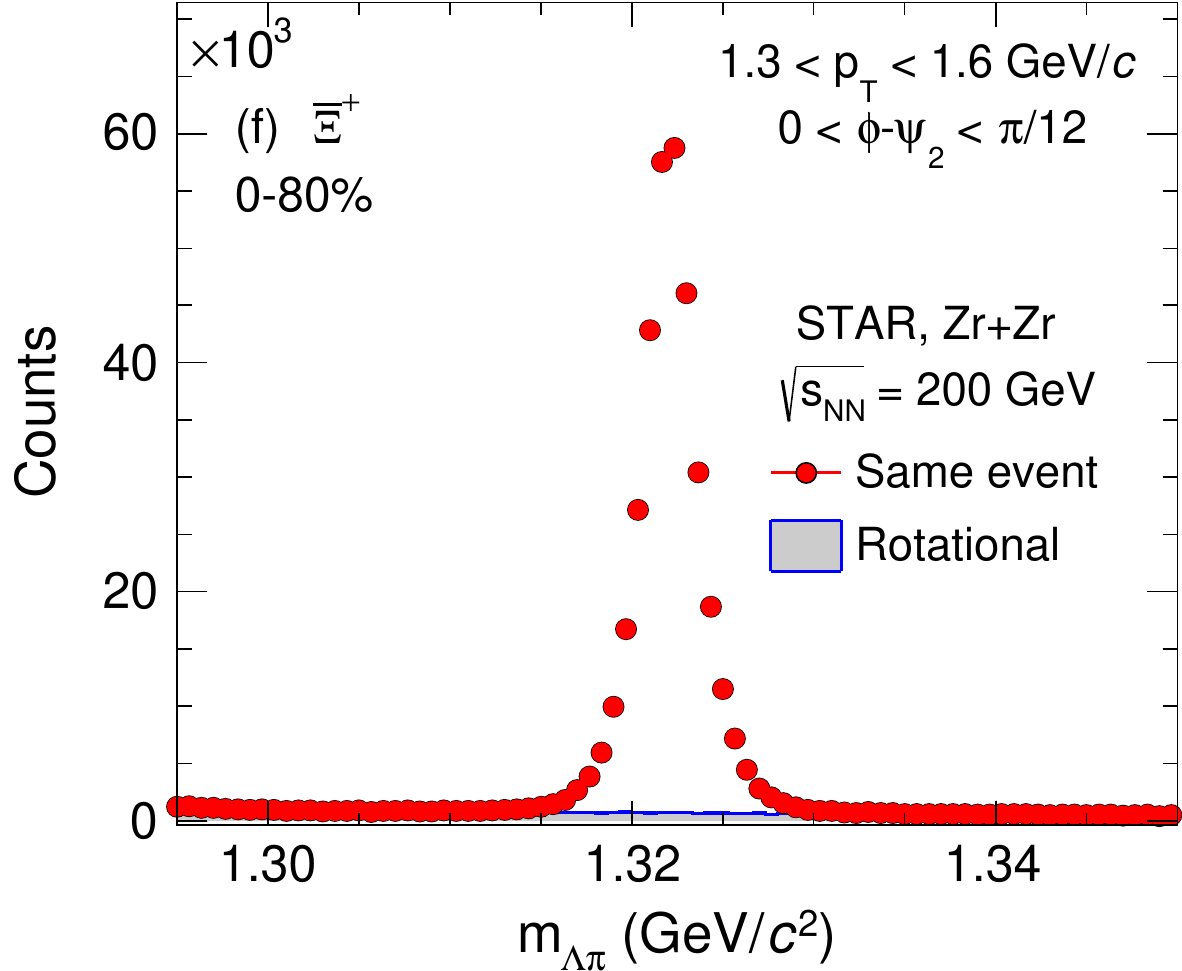}
\includegraphics[width=0.24\textwidth]{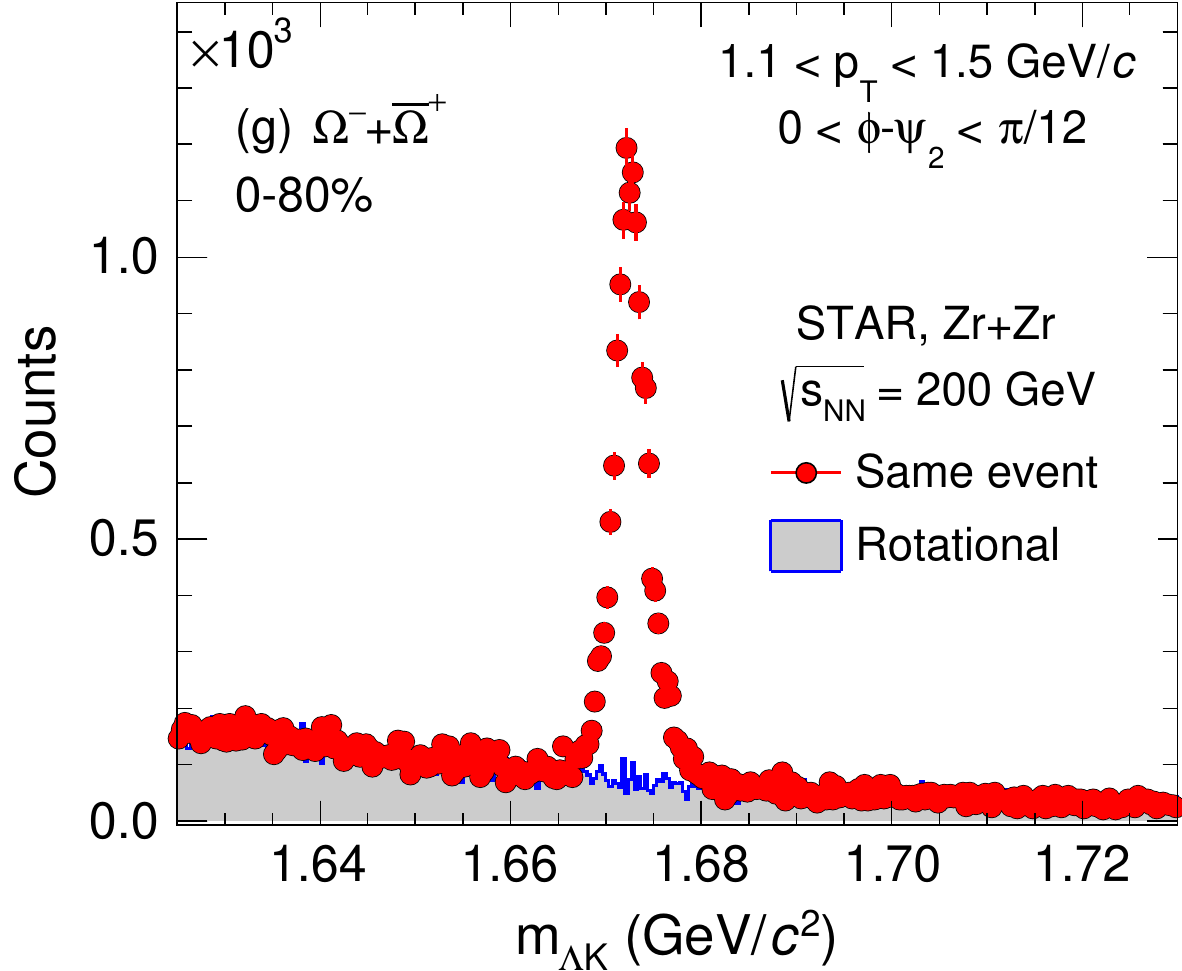}
\caption{Invariant mass distributions for (a) $K^{0}_{s}$, (b) $\phi$, (c) $\Lambda$, (d) $\overline{\Lambda}$, (e) $\Xi^{-}$, (f) $\overline{\Xi}^{+}$, and (g) $\Omega^{-}$+$\overline{\Omega}^{+}$ in minimum bias Zr+Zr collisions at $\sqrt{s_{NN}}$ = 200 GeV. The gray bands represent the combinatorial background. Rotational background technique is used for $K^{0}_{s}$, $\Lambda$, $\Xi$, and $\Omega$, while mixed event technique is used for $\phi$ mesons. Error bars represent the statistical uncertainties.}
\label{fig:invmass2}
\end{figure*}

The strange and multi-strange hadrons within mid-rapidity, $|y| <$ 1.0, are used for this analysis. The invariant mass distributions for (a) $K^{0}_{s}$, (b) $\phi$, (c) $\Lambda$, (d) $\overline{\Lambda}$, (e) $\Xi$, (f) $\overline{\Xi}$, and (g) $\Omega + \overline{\Omega}$ for a given $p_{\text{T}}$ range in minimum bias isobar collisions at $\snn$ are shown in Figs.~\ref{fig:invmass1} and~\ref{fig:invmass2}. The measured invariant mass distributions contain both signal (S) and background (B). A clear signal peak above the combinatorial background is seen around the rest mass of the particle. The combinatorial background shown as grey bands describes the shape of the background well.

\subsection{Flow Analysis Method}
\label{flow_method}
The elliptic flow is measured using the $\eta$ sub-event plane method~\cite{flowm2}. This method involves calculating the event plane, which is an estimation of the true reaction plane, using the azimuthal angle distribution of charged particle tracks. The azimuthal angle of the event plane, denoted as $\psi_{n}$, is calculated using the equation
\begin{equation}
\psi_{n} =\frac{1}{n}\tan^{-1} \left[\frac{\Sigma_{i} w_{i}\sin(n\phi_{i})}{\Sigma_{i} w_{i}\cos(n\phi_{i})}\right],
\label{eq:psi2}
\end{equation}
where n denotes a particular order of harmonic, $\phi_{i}$ is the azimuthal angle of $i^{th}$ particle, and $w_{i}$ is its weight. The weights used are the $\pt$ of the tracks. The second-order event plane angle is denoted by $\psi_{2}$. Various criteria are applied to select charged particle tracks to reconstruct the event plane angle. Specifically, tracks with DCA to the primary vertex of less than 2 cm and 15 or more hit points in the TPC detector are used to reconstruct the event plane angle. Additionally, only particles within $|\eta| <$ 1.0 and $\pt$ range between 0.2 to 2 GeV/$\it{c}$ from the TPC detector are used. The $\eta$ sub-event plane method involves event plane reconstruction in two $\eta$ windows (A: $-1.0 < \eta < -0.05$ and B: $0.05 < \eta < 1.0$), with a gap of $\Delta\eta =$ 0.1 between the two sub-events to further reduce non-flow effects primarily caused by resonance decays and (mini-)jet correlations. The non-uniform distribution of $\psi_{2}$ due to the non-homogeneity of the detector acceptance in the azimuthal direction is corrected by applying three different methods: $\phi$-weighting, re-centering, and shifting of the event plane angle~\cite{flow1}.
\begin{figure}[!htbp]
\centering
\includegraphics[width=0.43\textwidth]{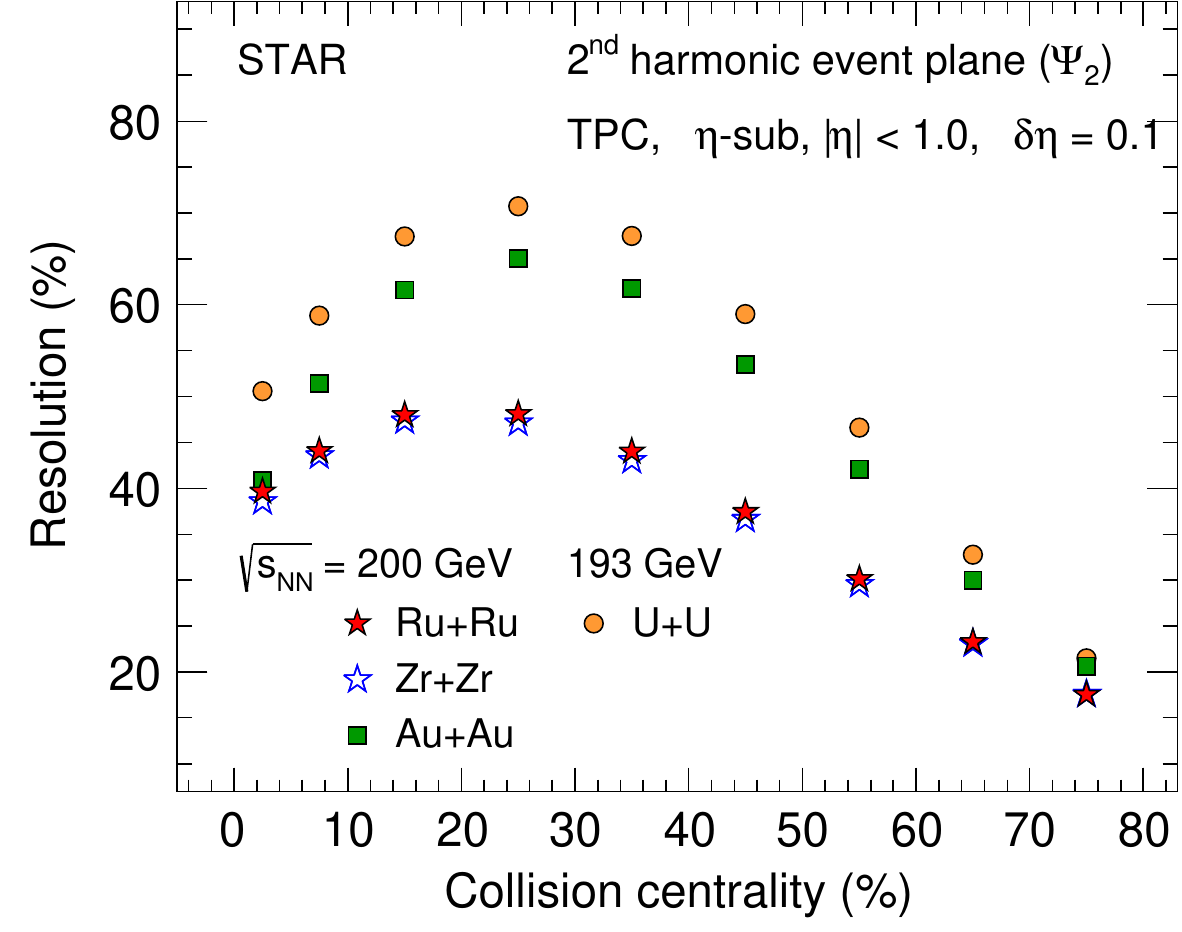}
\caption{\small{Second-order $\eta$-sub event plane angle ($\psi_{2}$) resolution obtained from the TPC as a function of centrality in Ru+Ru and Zr+Zr collisions at $\sqrt{s_{NN}}$ = 200 GeV. The event plane resolution is compared with Au+Au collisions at $\sqrt{s_{NN}}$ = 200 GeV and U+U collisions at $\sqrt{s_{NN}}$ = 193 GeV~\cite{lflow,STARUU}.}}
\label{fig:epres}
\end{figure}

The elliptic flow is calculated with respect to the second-order event plane angle as $v_{2} = \langle \cos\left[2(\phi-\psi_{2})\right] \rangle$. The angular brackets indicate an average over all particles and events in a given phase space. This calculated $\vtwo$ value is then corrected for the event-plane resolution. The $\eta$ sub-event plane resolution (R) with the assumption of only flow correlations between the two sub-events is defined as~\cite{flowm2}
\begin{equation}
R = \sqrt{\left\langle\cos\left[2\left(\psi_{2}^{A} - \psi_{2}^{B}\right)\right]\right\rangle}.  
\label{eq:etasubres}
\end{equation}
In this analysis, event plane resolution is calculated for nine different centrality classes using similar technique as for other collision systems~\cite{lflow,STARUU}. Figure~\ref{fig:epres} shows the second-order $\eta$ sub-event plane resolution as a function of centrality in isobar collisions at $\snn$. The event plane resolution is comparable between the two isobars Zr+Zr and Ru + Ru and is strongly dependent on the collision centrality. The resolution is compared with that in Au+Au collisions at $\sqrt{s_{NN}}$ = 200 GeV and U+U collisions at $\sqrt{s_{NN}}$ = 193 GeV showing similar a centrality dependence. The comparison shows a lower resolution in the isobar collisions owing to the smaller number of participating nucleons. 

\begin{figure*}[!htbp]
\centering
\includegraphics[width=0.24\textwidth]{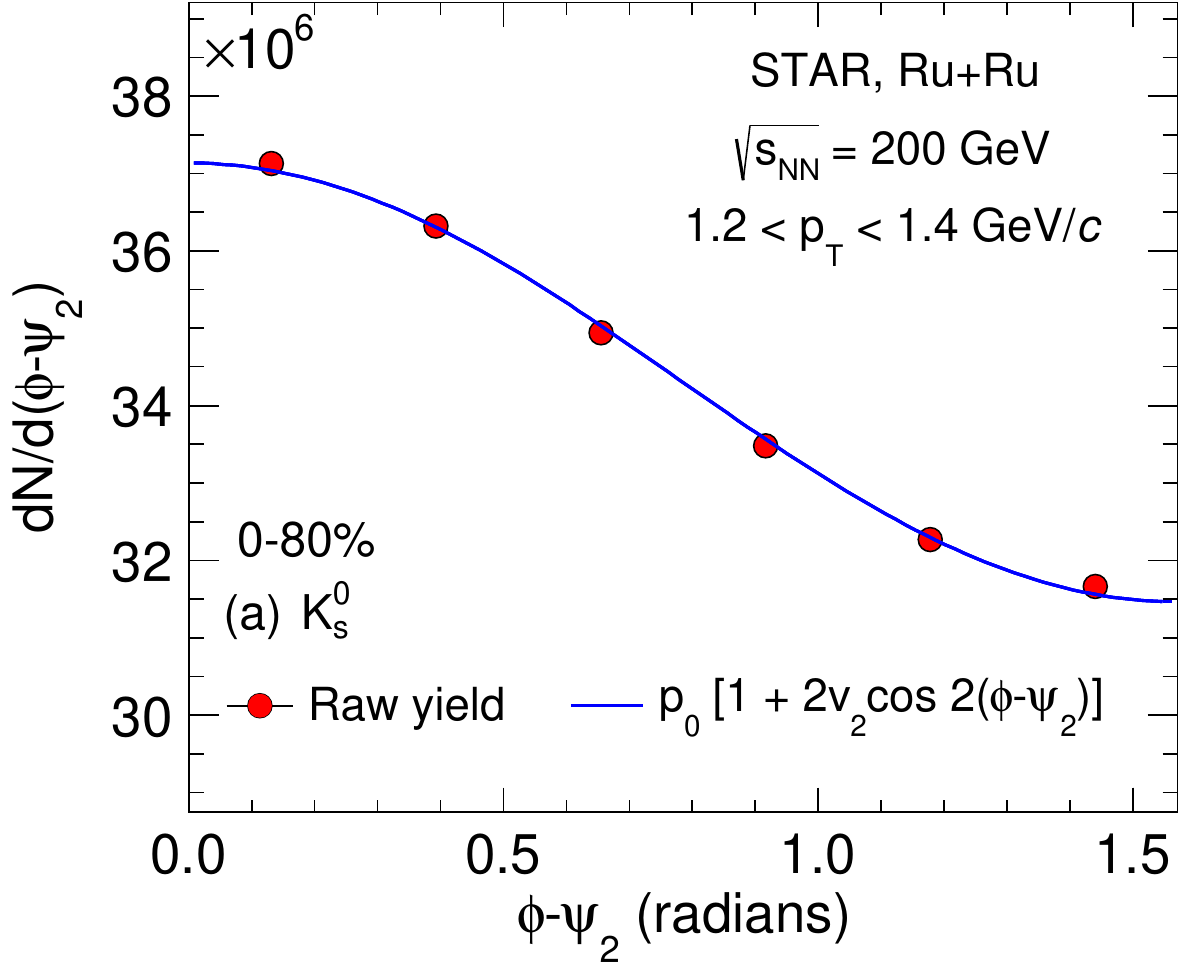} 
\includegraphics[width=0.24\textwidth]{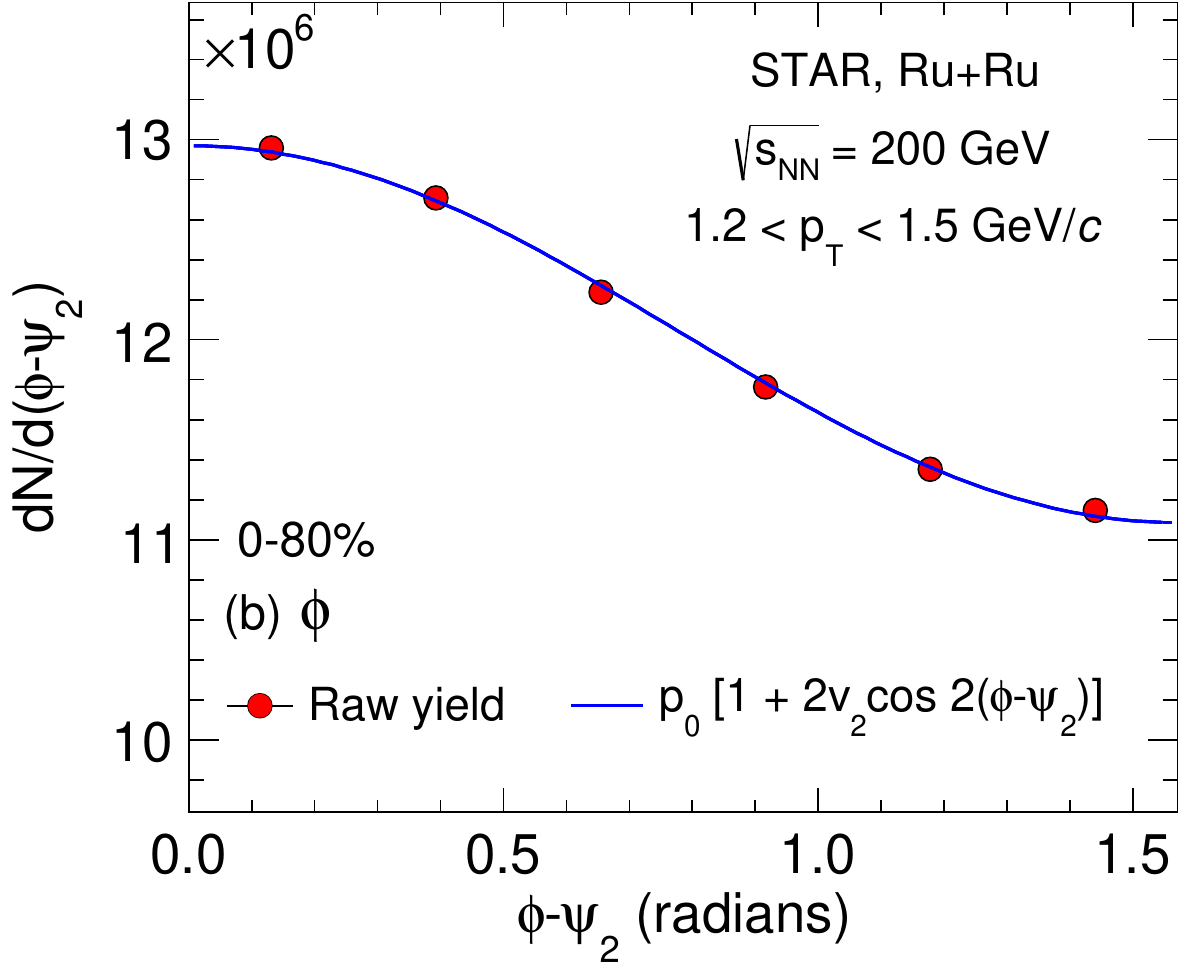}
\includegraphics[width=0.24\textwidth]{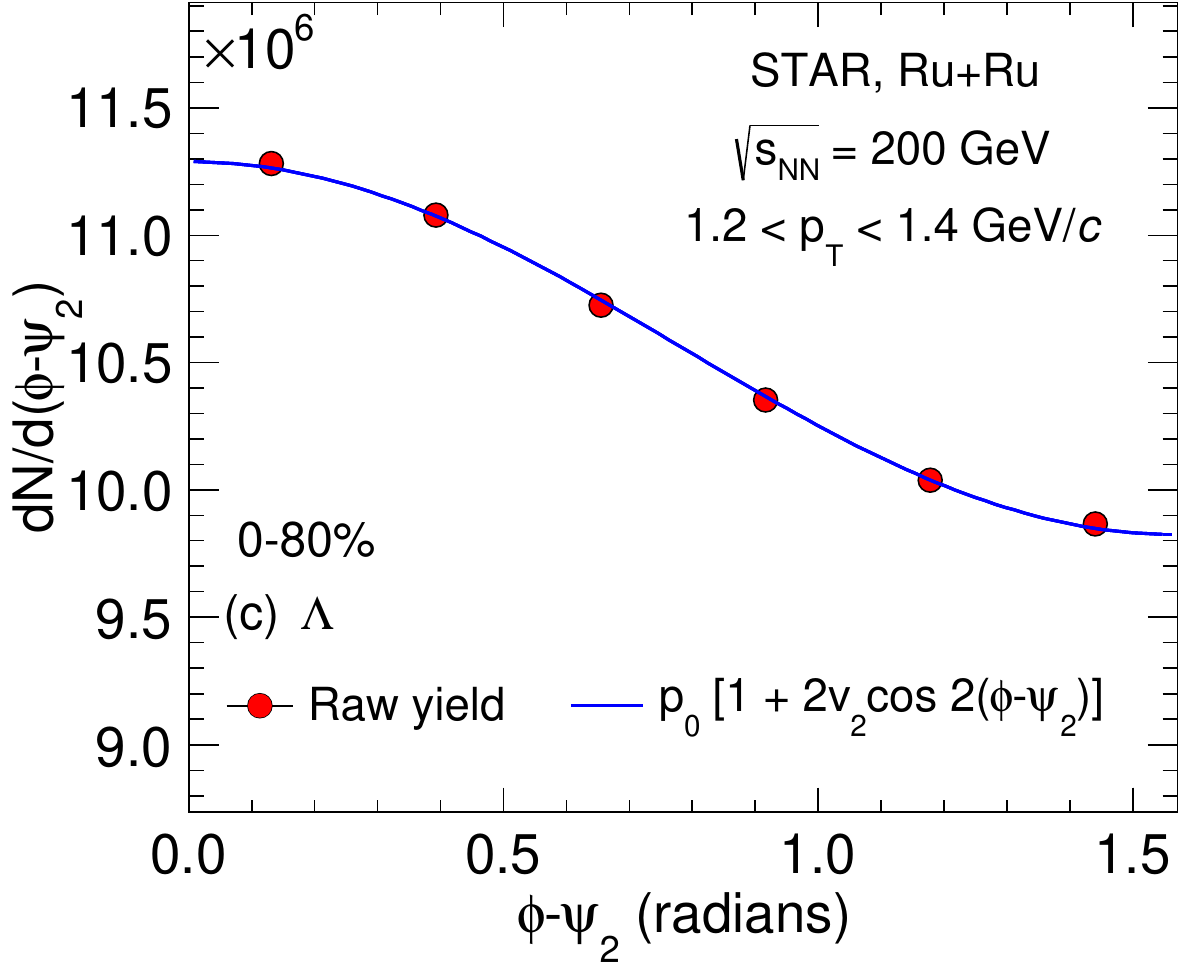}
\includegraphics[width=0.24\textwidth]{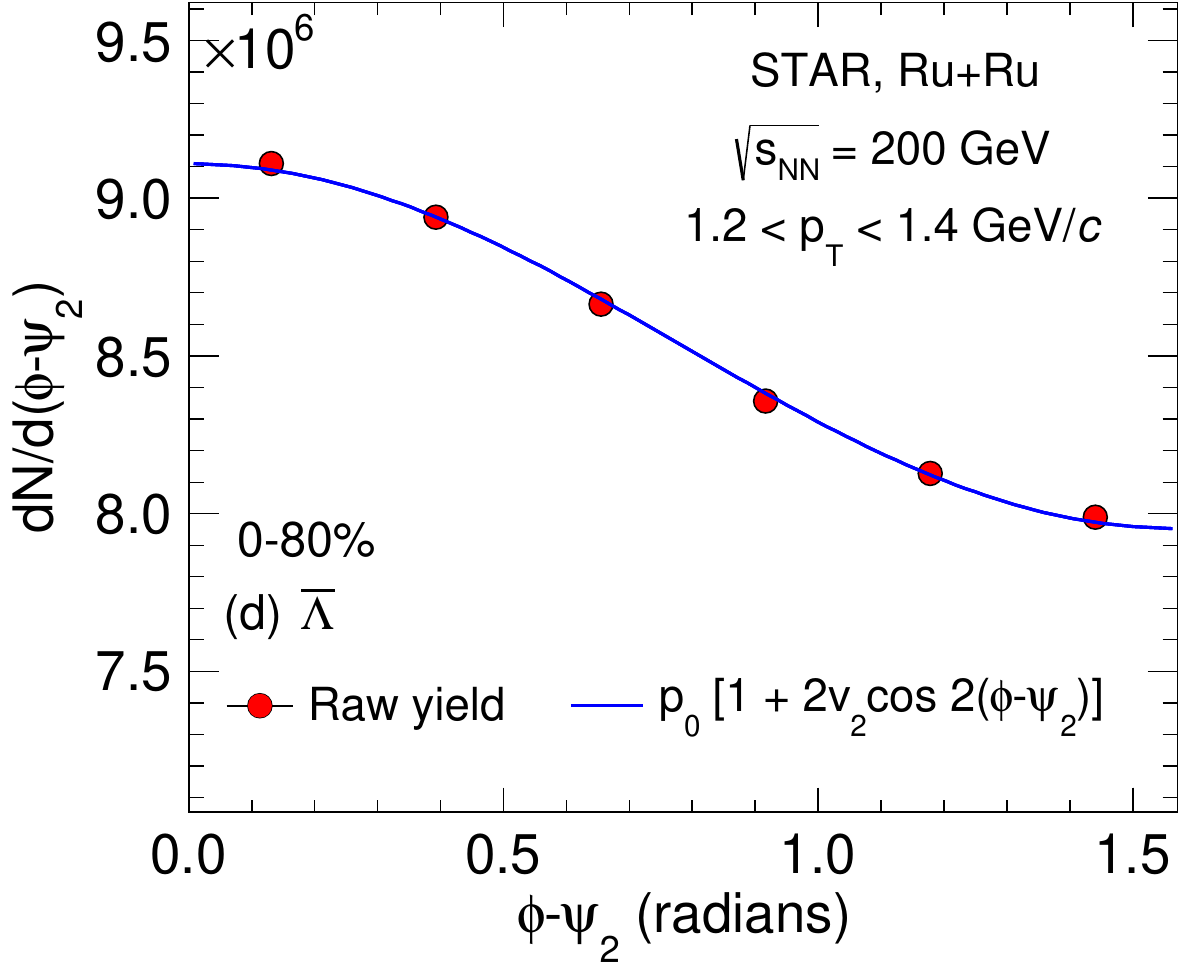} 
\includegraphics[width=0.24\textwidth]{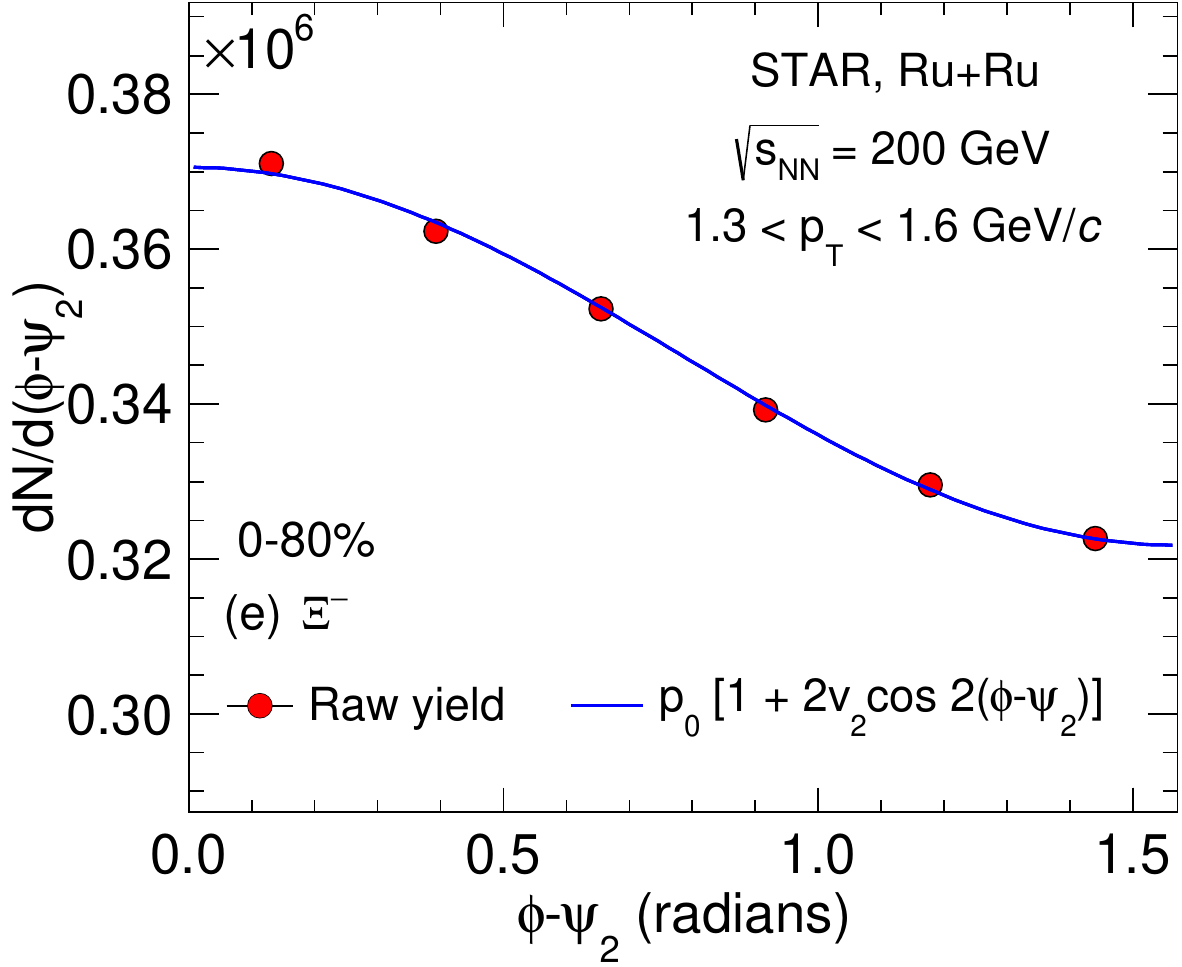}
\includegraphics[width=0.24\textwidth]{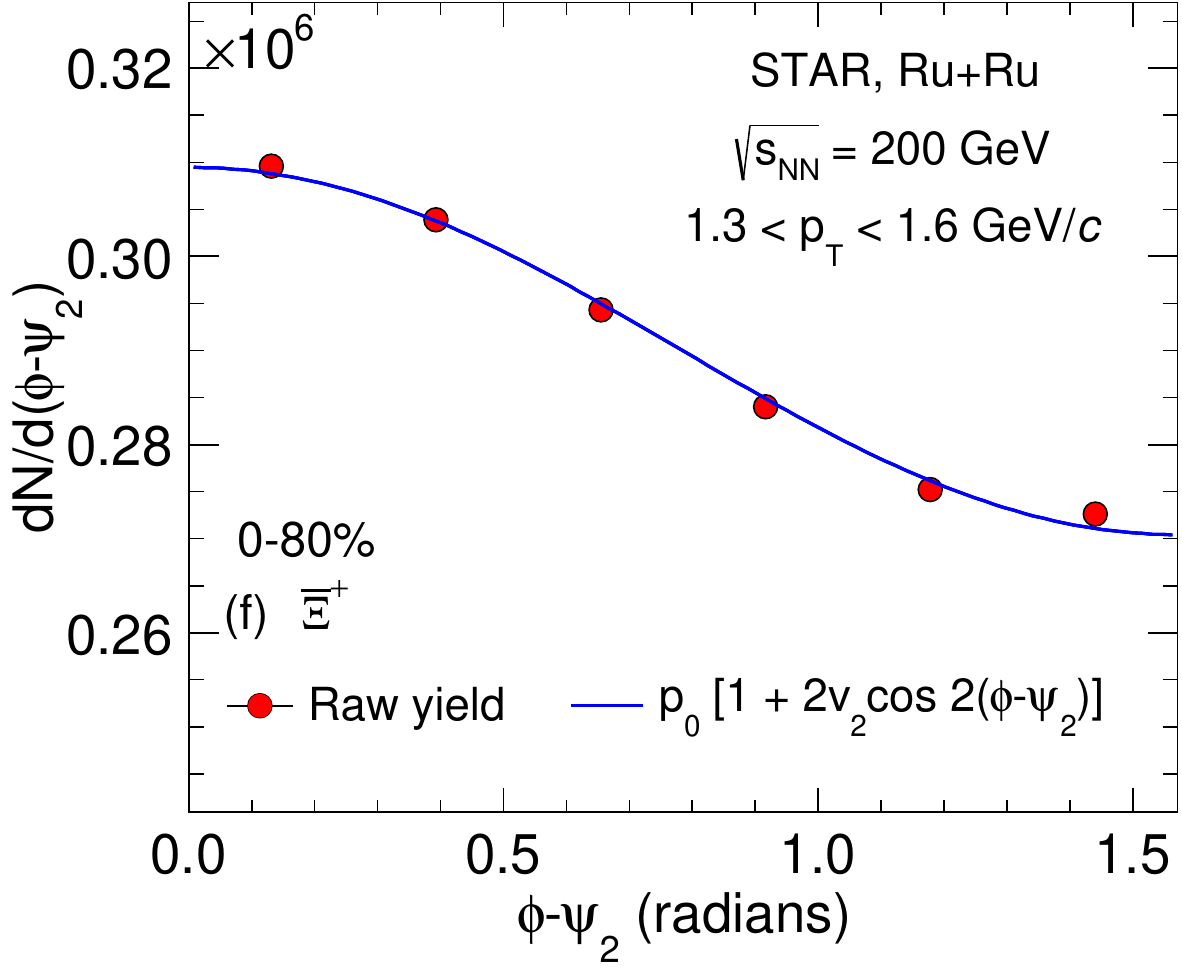}
\includegraphics[width=0.24\textwidth]{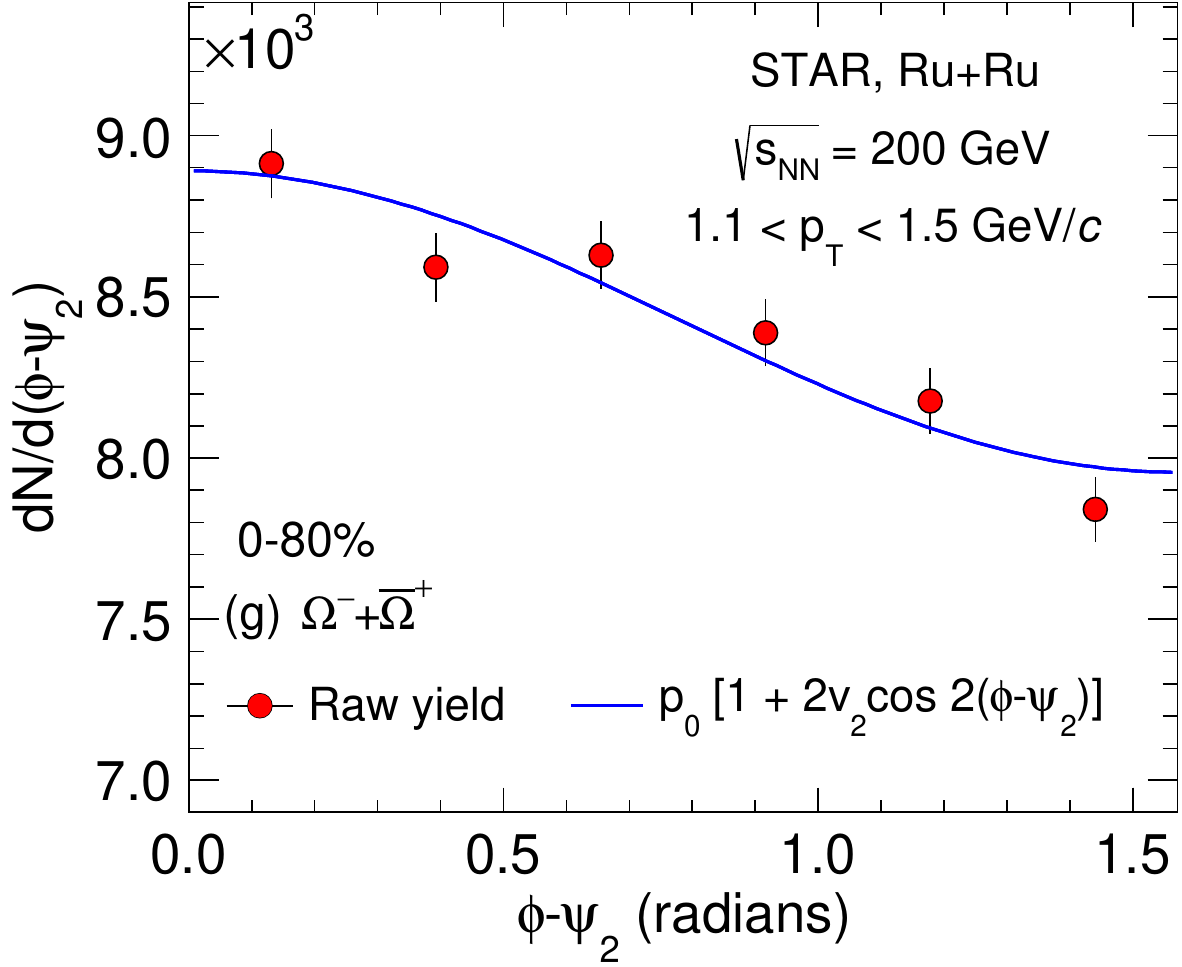}
\caption{Raw yield as a function of $\phi-\psi_{2}$ for (a) $K^{0}_{s}$, (b) $\phi$, (c) $\Lambda$, (d) $\overline{\Lambda}$, (e) $\Xi^{-}$, (f) $\overline{\Xi}^{+}$, and (g) $\Omega^{-}$+$\overline{\Omega}^{+}$ at mid-rapidity ($|y| <$ 1.0) in minimum bias Ru+Ru collisions at $\sqrt{s_{NN}}$ = 200 GeV. The solid lines represent fits to the data to extract $v_{2}$. Error bars represent the statistical uncertainties.}
\label{fig:vnfit1}
\end{figure*}

\begin{figure*}[!htbp]
\centering
\includegraphics[width=0.24\textwidth]{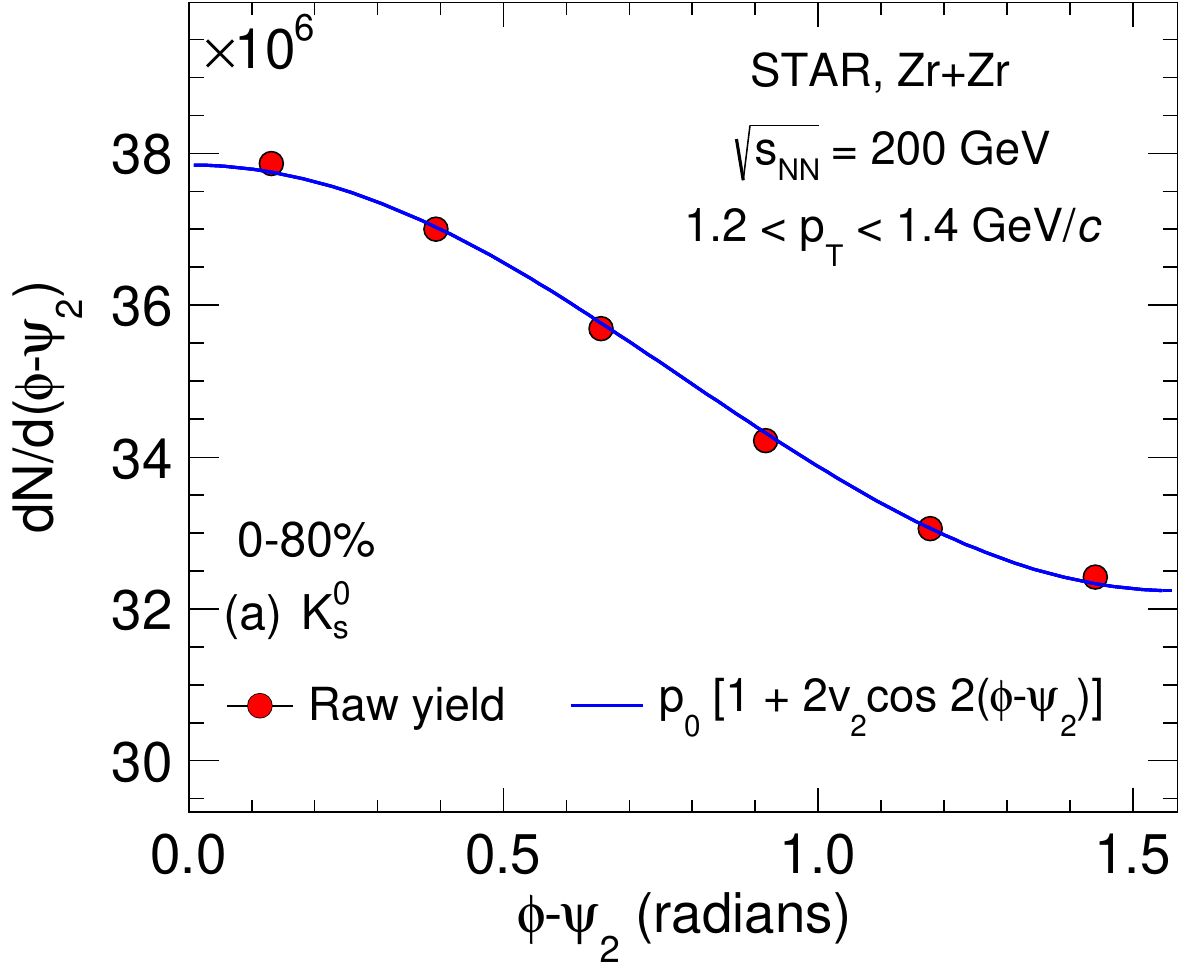} 
\includegraphics[width=0.24\textwidth]{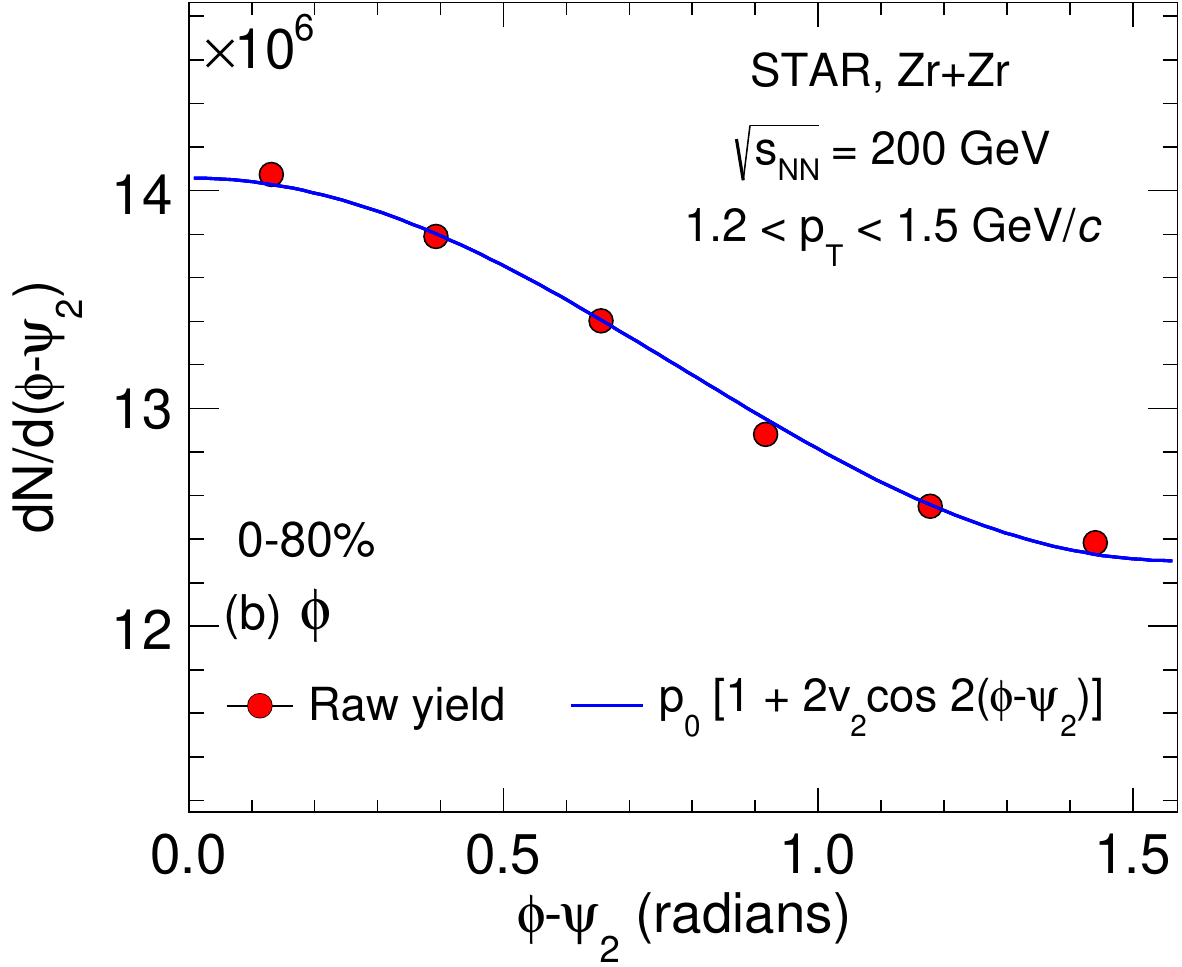}
\includegraphics[width=0.24\textwidth]{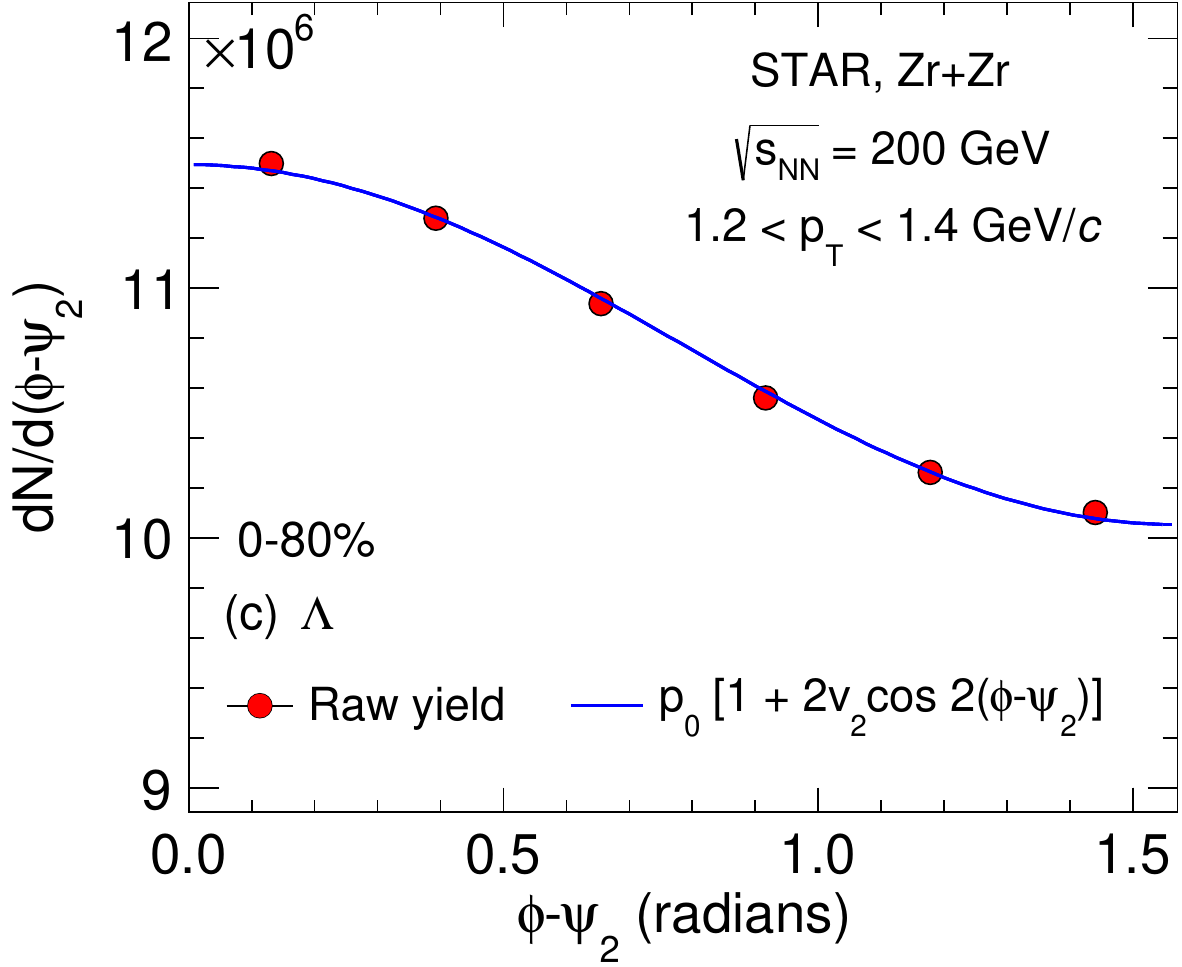}
\includegraphics[width=0.24\textwidth]{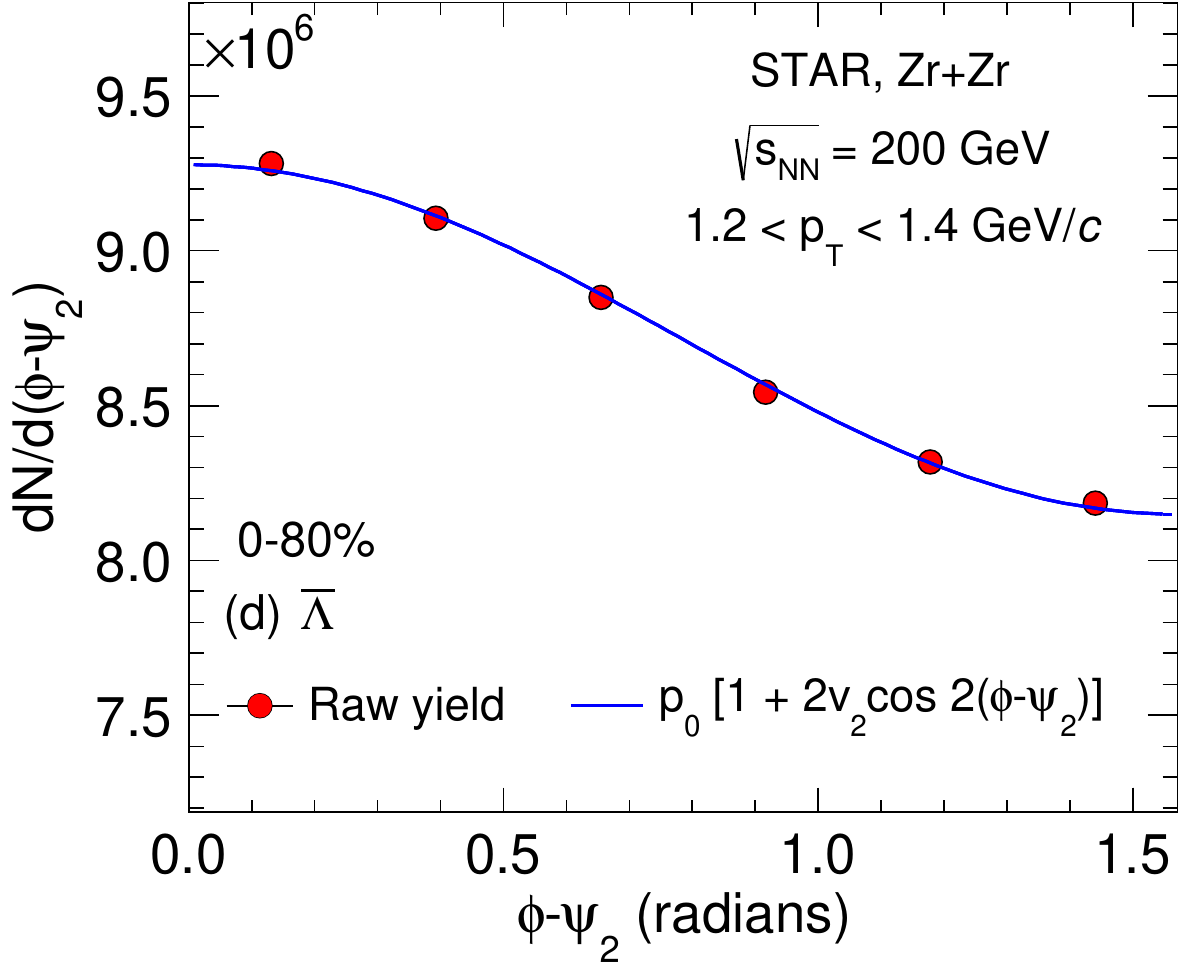} 
\includegraphics[width=0.24\textwidth]{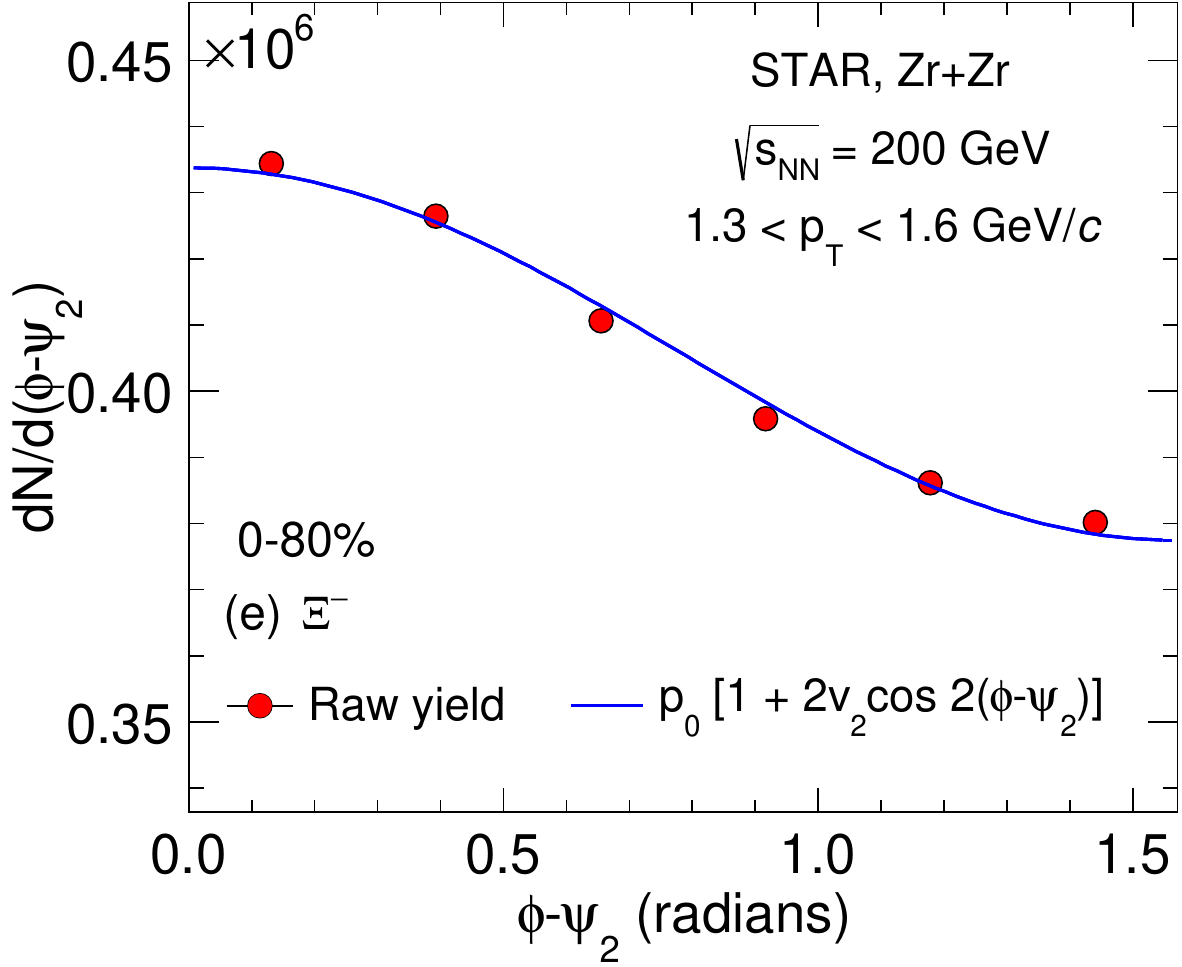}
\includegraphics[width=0.24\textwidth]{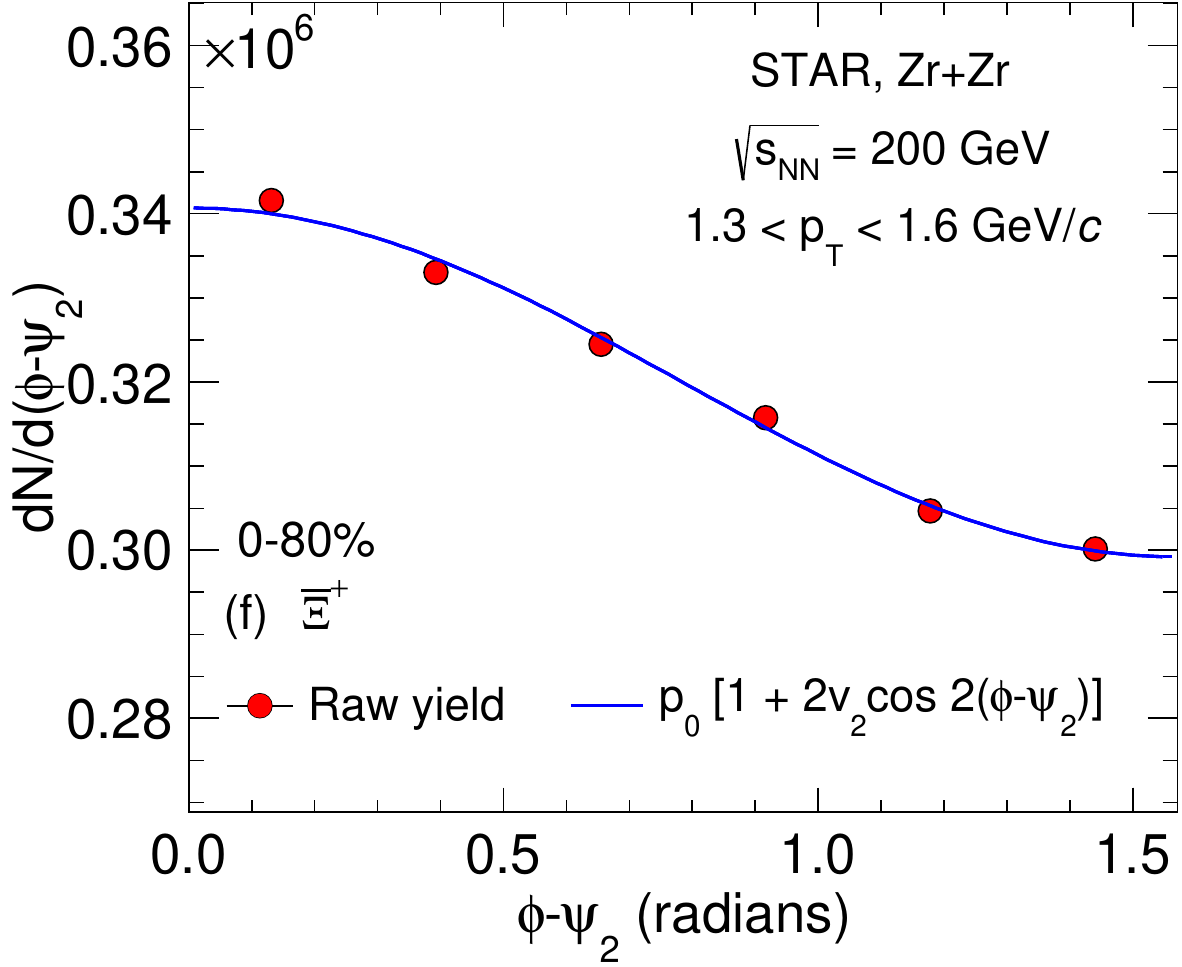}
\includegraphics[width=0.24\textwidth]{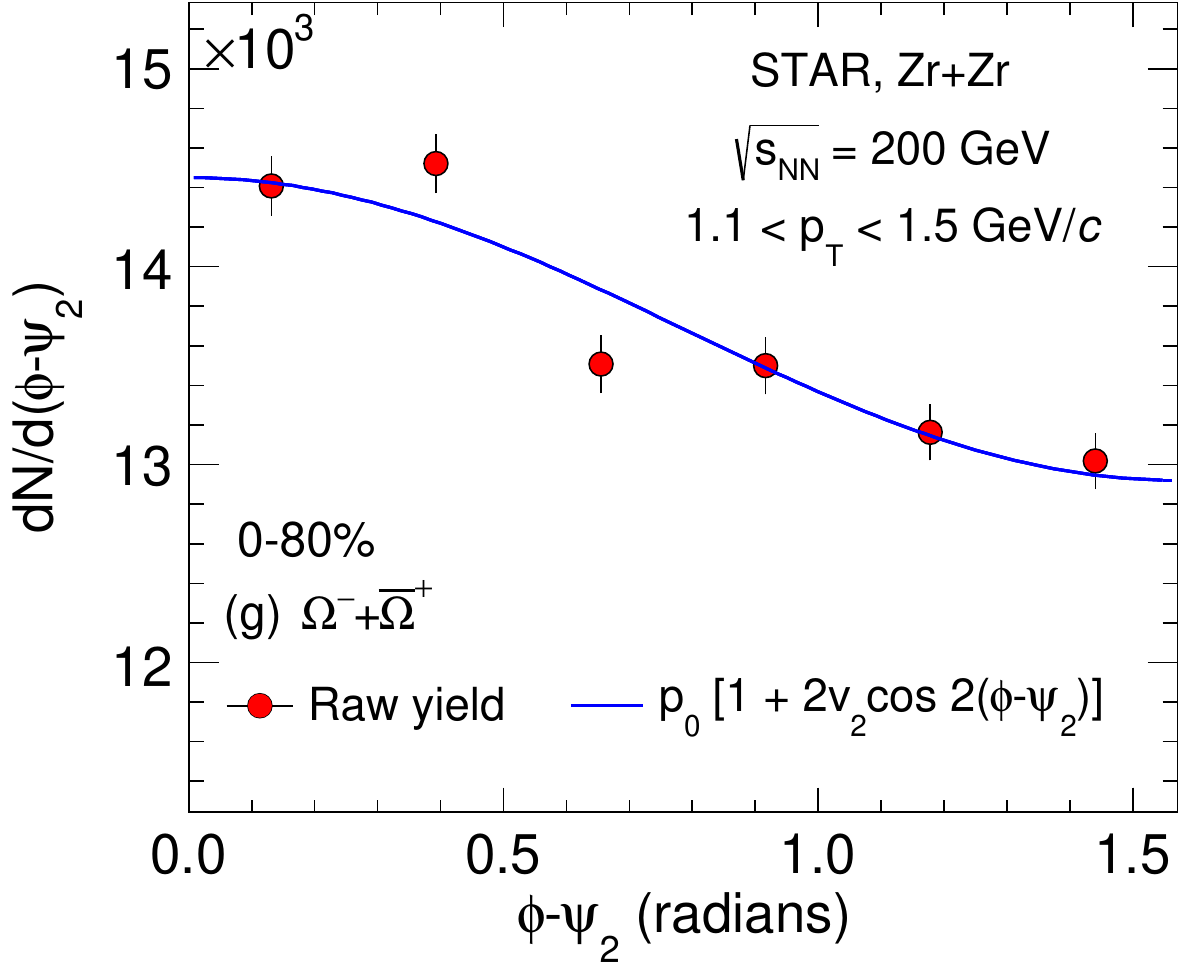}
\caption{Raw yield as a function of $\phi-\psi_{2}$ for (a) $K^{0}_{s}$, (b) $\phi$, (c) $\Lambda$, (d) $\overline{\Lambda}$, (e) $\Xi^{-}$, (f) $\overline{\Xi}^{+}$, and (g) $\Omega^{-}$+$\overline{\Omega}^{+}$ at mid-rapidity ($|y| <$ 1.0) in minimum bias Zr+Zr collisions at $\sqrt{s_{NN}}$ = 200 GeV. The solid lines represent fits to the data to extract $v_{2}$. Error bars represent the statistical uncertainties.}
\label{fig:vnfit2}
\end{figure*}

The raw yield of the $\phi$ mesons is obtained through a Breit-Wigner and second-order polynomial fit to the invariant mass distribution. For $K_{s}^{0}$, $\Lambda$, $\Xi$, and $\Omega$, a bin-counting approach is used. The raw yield is obtained in different $\pt$ intervals as a function of $\phi-\psi_{2}$ angle in each of the centrality classes~\cite{ksl1,ksl2}. The observed $v_{2}$ is obtained by fitting particle yields as a function of $\phi-\psi_{2}$ with the functional form given by the equation 
\begin{equation}
\frac{dN}{d(\phi-\psi_{2})} = p_{0} \left[ 1+ 2 v_{2} \cos 2(\phi-\psi_{2}) \right],
\label{eq:vnfunction}
\end{equation}
where $p_{0}$ is a normalization parameter. Figures~\ref{fig:vnfit1} and~\ref{fig:vnfit2} show particle yields as a function of $\phi-\psi_{2}$ for a given $\pt$ interval in minimum bias Ru+Ru and Zr+Zr collisions at $\snn$. The analysis is performed for various combined centrality classes (0-10\%, 10-40\%, and 40-80\%) for which the corresponding resolution is evaluated by taking the particle-yield-weighted average of resolution in the finer centrality classes. Thereafter, the true $v_{2}$ is obtained by dividing the observed $v_{2}$ with the corresponding event plane resolution. 

\section{Systematic uncertainties}
\label{sys_error}
The systematic uncertainties are evaluated by varying the cuts for event selection, track selection, particle identification, and topological selection. We vary the lower $V_{z}$ cut from -35 cm to -30 cm and the higher $V_{z}$ cut from 25 cm to 20 cm to evaluate systematic uncertainty in event selection. The particle identification cut, such as the $N_{fit}^{hits}$ and $n_{\sigma}$ values, are varied by $\sim$10-20\% from the default value. In order to obtain the systematic uncertainty from decay topology, we vary DCA of the parent particle, the DCA between daughters, the DCA of the daughters to the primary vertex, and the decay length of the particle by $\sim$20\% from the default values. We also estimate systematic uncertainty associated with the background estimation by varying the order of the polynomial fit function from second to third order. 

For each systematic variation, a quantity, $|\Delta|$, defined as the magnitude of the difference of $v_{2}$ between the default cut and a particular variation in the cut, is calculated. Then, the statistical fluctuation in the $\vtwo$ value is quantified by
\begin{equation}
\sigma_{\Delta} = \sqrt{ |\sigma_{def}^{2} - \sigma_{i}^{2} |}, 
\end{equation} where $\sigma_{def}$ and $\sigma_{i}$ are the statistical uncertainties associated with the $\vtwo$ values using the default and a particular cut~\cite{Barlow}. If $\sigma_{\Delta}$ is greater than $|\Delta|$, then it implies that the deviation in $\vtwo$ value for the corresponding cut variation from default value is within the statistical uncertainty. In that case, a systematic uncertainty of zero is assigned. Otherwise, the systematic uncertainty is calculated as
\begin{equation}
\sigma_{sys} = \sqrt{\Delta^{2} - \sigma_{\Delta}^{2}}.
\end{equation}
This procedure is applied for each given source of systematic uncertainty. The total systematic uncertainty on the $\vtwo$ value is obtained by adding all the sources of systematic uncertainties in quadrature. Table~\ref{tab:syserr} gives the relative systematic uncertainties for all the particle species in isobar collisions at $\snn$. 
\begin{table*}[!htbp]
\centering
\caption{Relative systematic uncertainties on $\pt$-differential $\vtwo$ in isobar collisions at $\snn$. All numbers represent percent uncertainties.}
\vspace{0.2cm}
\label{tab:syserr}
\begin{tabular}{ccccccccc}
\hline
\hline
 & \multicolumn{4}{c}{Ru} & \multicolumn{4}{c}{Zr} \\ [0.2ex]
\hline
Particle & 0-80\% & 0-10\% & 10-40\% & 40-80\% & 0-80\% & 0-10\% & 10-40\% & 40-80\% \\[0.2ex]
 \hline
$K_{s}^{0}$ & 0.3 & 0.9 & 0.4 & 0.4 & 0.3 & 1.1 & 0.4 & 0.4 \\
$\Lambda$ & 2 & 5 & 2 & 8 & 2 & 5 & 3 & 1  \\
$\overline{\Lambda}$ & 2 & 7 & 0.7 & 1 & 1 & 3 & 2 & 1 \\
$\Xi^{-}$ & 3 & 3 & 1 & 1 & 1 & 3 & 1 & 2   \\
$\overline{\Xi}^{+}$& 1 & 2 & 1 & 2 & 1 & 3 & 2 & 1  \\
$\Omega^- + \overline{\Omega}^{+}$ & 7 & 19 & 13 & 12 & 5 & 12 & 4 & 14 \\
$\phi$ & 7 & 13 & 5 & 7 & 8 & 12 & 2 & 5 \\
\hline
\hline
\end{tabular}
\end{table*}

\section{Results and Discussion}
\label{result}
\subsection{$p_{\text{T}}$ dependence of $v_{2}$}
\label{ptdep}
\begin{figure*}[!htbp]
\centering
\includegraphics[totalheight=5cm]{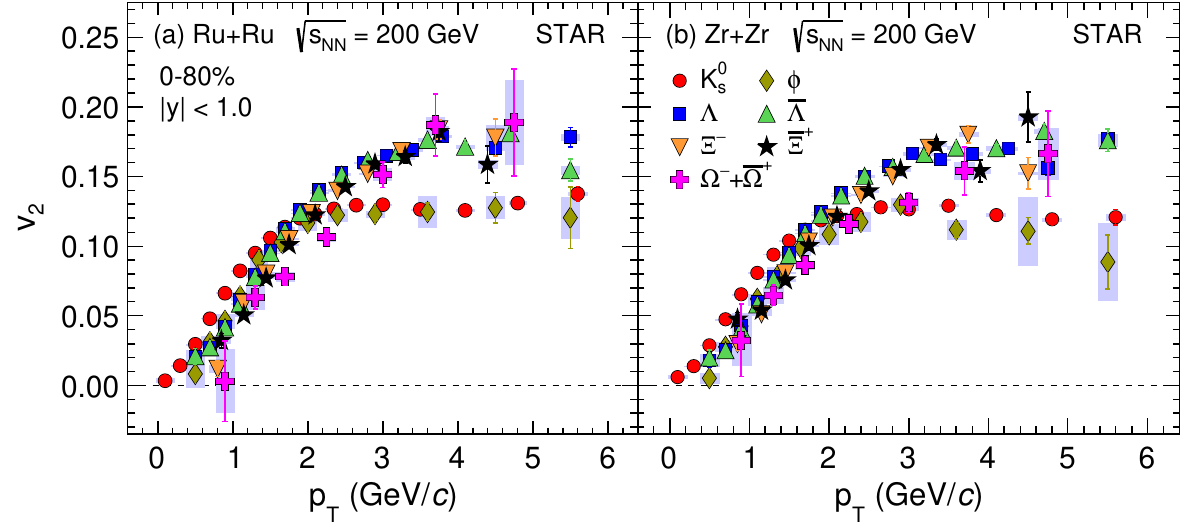}
\caption{\small{$v_{2}$ as a function of $\pt$ at mid-rapidity ($|y| < 1.0$) in minimum bias (a) Ru+Ru and (b) Zr+Zr collisions at $\snn$. The error bars represent statistical uncertainties. The bands represent point-by-point systematic uncertainties.}}
\label{fig:massdep}
\end{figure*}

In this section, the $p_{\text{T}}$ dependence of $v_{2}$ is reported for strange and multi-strange hadrons at mid-rapidity ($|y| <$ 1.0) in isobar collisions at $\snn$. Figure~\ref{fig:massdep} shows the particle species dependence of $v_{2}$($\pt$) in minimum bias (0-80\%) Ru+Ru and Zr+Zr collisions. The $\vtwo$ at low $\pt$ shows higher values for lighter hadrons like $K_s^0$, implying hydrodynamic behaviour in isobaric collisions similar to that observed in Au+Au collisions at $\snn$~\cite{AuCent}. At intermediate $\pt$ ($\sim$2-5 GeV/c), the hydrodynamic mass ordering breaks, and the $\vtwo$($\pt$) shows splitting between baryons and mesons in both Ru+Ru and Zr+Zr collisions. This splitting implies the dominance of partonic degrees of freedom in the intermediate $\pt$ range, which can be explored through a number of constituent quarks ($n_{q}$) scaling of $\vtwo$.

\begin{figure*}[!htbp]
\centering
\includegraphics[totalheight=6.5cm]{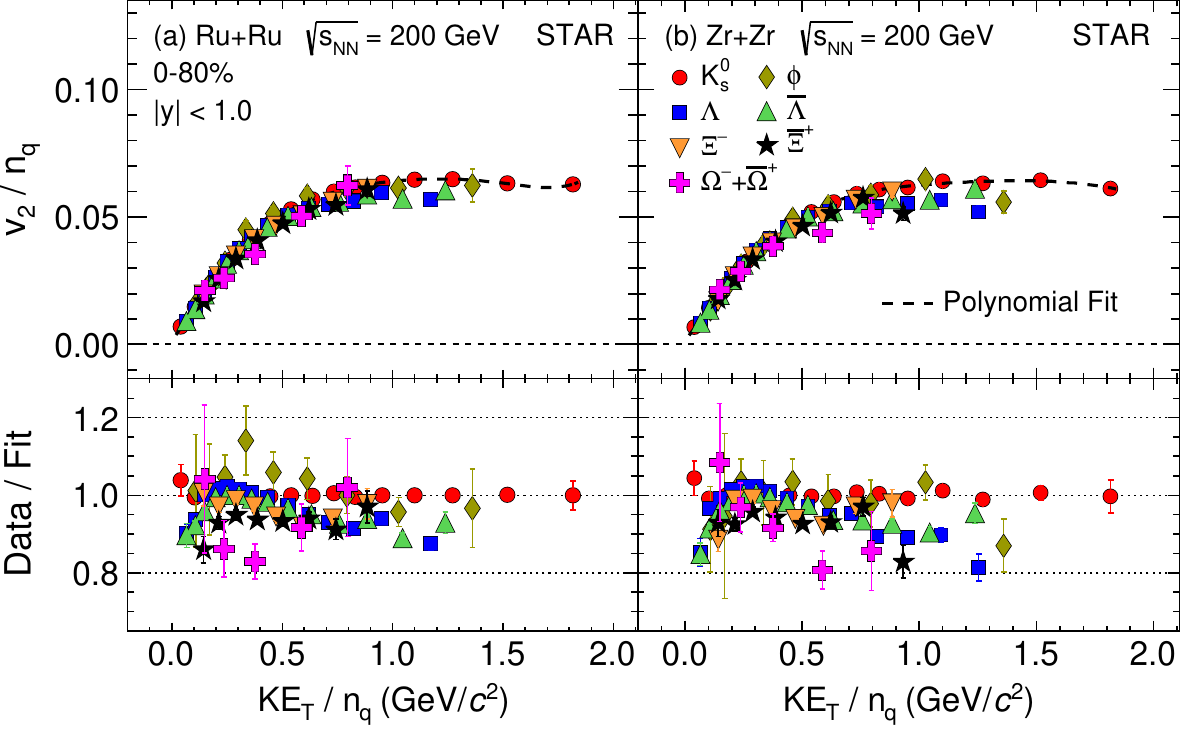}
\caption{\small{Flow coefficient $v_{2}$ as a function of transverse kinetic energy ($\kt$) for (multi-)strange hadrons at mid-rapidity for 0-80\% centrality in Ru+Ru and Zr+Zr collisions at $\snn$, scaled by number of constituent quarks ($n_{q}$). Ratio of $\vtwo$ with respect to the polynomial fit of $K_{s}^{0}$ $\vtwo$ are shown in the corresponding lower panels. Error bars represent statistical and systematic uncertainties added in quadrature.}}
\label{fig:ncq}
\end{figure*}

Figure~\ref{fig:ncq} shows the number of constituent quark scaled $v_{2}$ as a function of $\kt$/$n_{q}$ in Ru+Ru and Zr+Zr collisions at $\snn$. Here, $\kt$ represents transverse kinetic energy $\kt$ = $m_{T} - m_{0}$, where $m_{T}$ = $\sqrt{\pt^{2} + m_{0}^{2}}$ and $m_{0}$ is the rest mass of the particle. The particle mass dependence of $\vtwo$ at low $\pt$ is removed by subtracting the rest mass contribution from $\pt$. A sixth-order polynomial fit is performed to the $\vtwo$ of $K_{s}^{0}$, then the ratio of the $\vtwo$ of other particles with respect to that fit function is taken and shown in the bottom panels of Fig.~\ref{fig:ncq}. All the studied (multi-)strange hadrons follow a number of constituent quark scaling to within 20\%. Specifically, the multi-strange hadrons ($\phi$ and $\Omega$) having lower interaction cross-section with other hadrons also exhibit $n_{q}$ scaling. This $n_{q}$ scaling suggests the formation of QGP medium with partons as dominant degrees of freedom and quark coalescence as the mechanism of particle production. Interestingly, the observation of $n_{q}$ scaling in the relatively smaller isobar collision system, similar to Au+Au and U+U, indicates the collective behaviour of the produced medium~\cite{idflowe1,STARUU}. 

\subsection{Centrality dependence of $v_{2}$}
\label{centdep}
\begin{figure*}[!htbp]
\centering
\includegraphics[totalheight=7.cm]{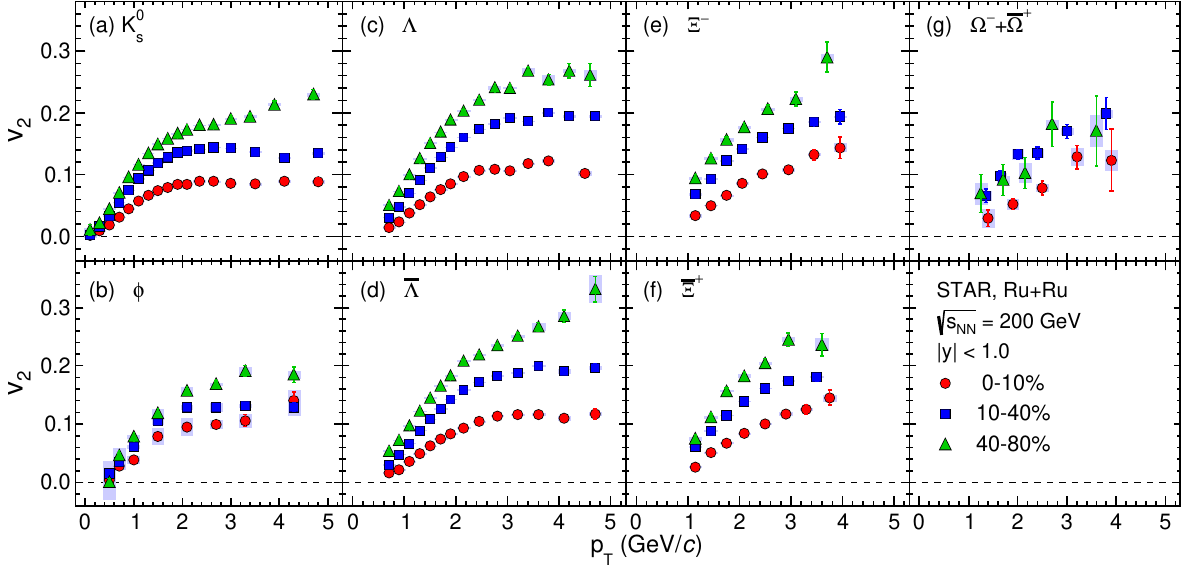}
\caption{\small{$v_{2}$ as a function of $\pt$ at mid-rapidity in Ru+Ru collisions at $\snn$ for centrality classes 0-10\%, 10-40\%, and 40-80\%. The error bars represent statistical uncertainties. The bands represent point-by-point systematic uncertainties.}}
\label{fig:rucentdep}
\end{figure*}

\begin{figure*}[!htbp]
\centering
\includegraphics[totalheight=7.cm]{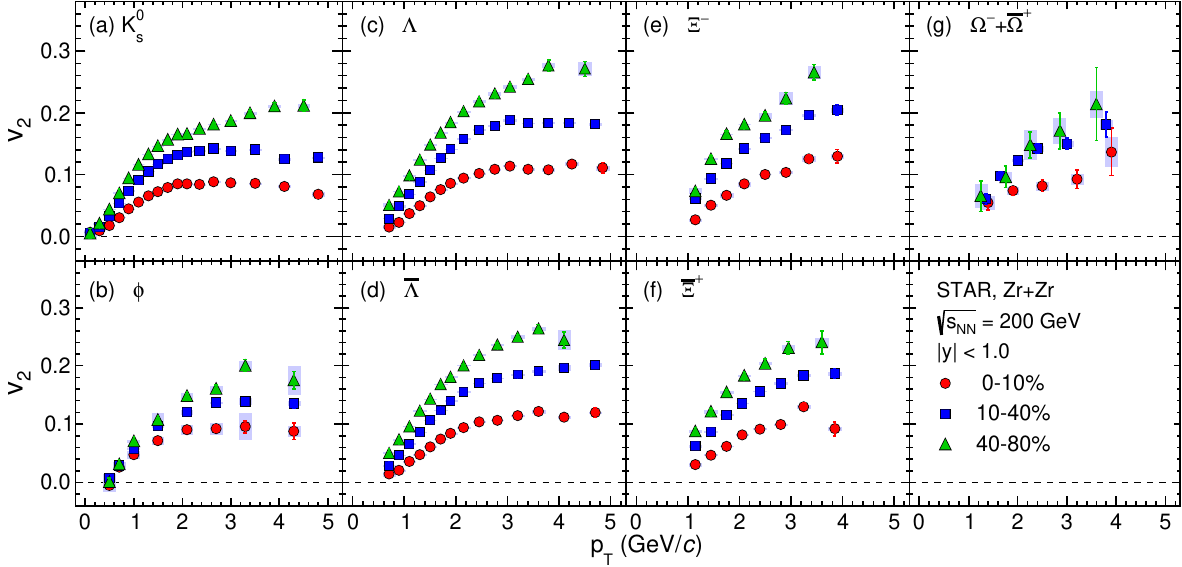}
\caption{\small{$v_{2}$ as a function of $\pt$ at mid-rapidity in Zr+Zr collisions at $\snn$ for centrality classes 0-10\%, 10-40\%, and 40-80\%. The error bars represent statistical uncertainties. The bands represent point-by-point systematic uncertainties.}}
\label{fig:zrcentdep}
\end{figure*}

Elliptic flow primarily originates from the geometric anisotropy of the initial overlap region between the colliding nuclei, which can be explored by studying the centrality dependence of $\vtwo$($\pt$). Figures~\ref{fig:rucentdep} and~\ref{fig:zrcentdep} show $\vtwo$ as a function of $\pt$ in different centrality classes from central (0-10\%), mid-central (10-40\%), and peripheral (40-80\%) collisions in Ru+Ru and Zr+Zr collisions, respectively. The magnitude of $\vtwo$ increases from central to peripheral collisions for all the studied hadrons in both the systems. This centrality dependence reflects the higher initial geometrical anisotropy in the peripheral collisions compared to central collisions. 

\begin{figure*}[!htbp]
\centering
\includegraphics[totalheight=6.cm]{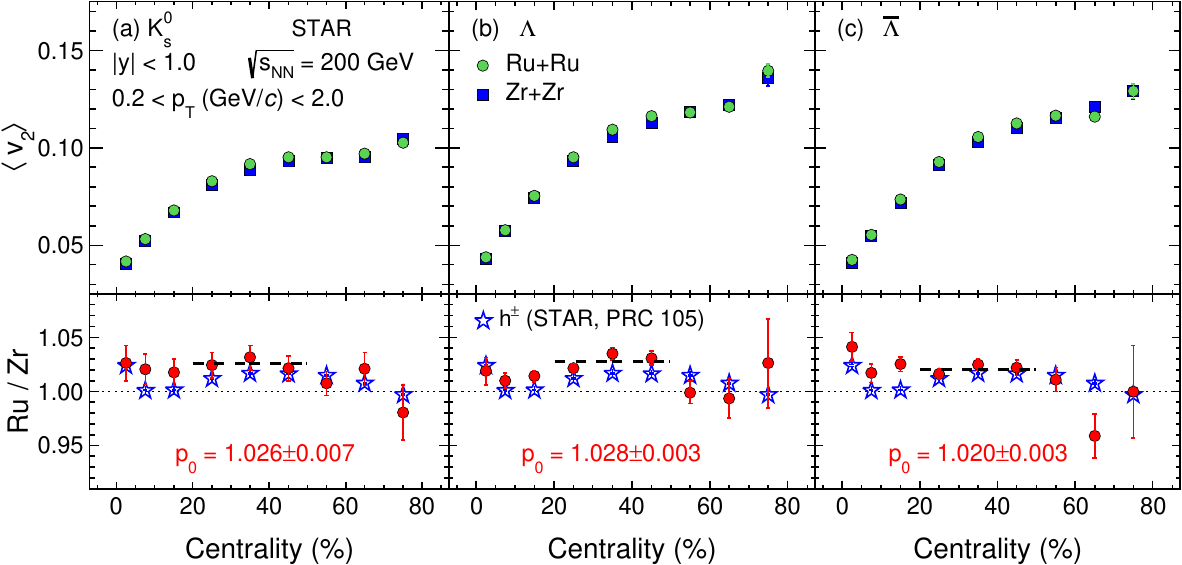}
\includegraphics[totalheight=6.cm]{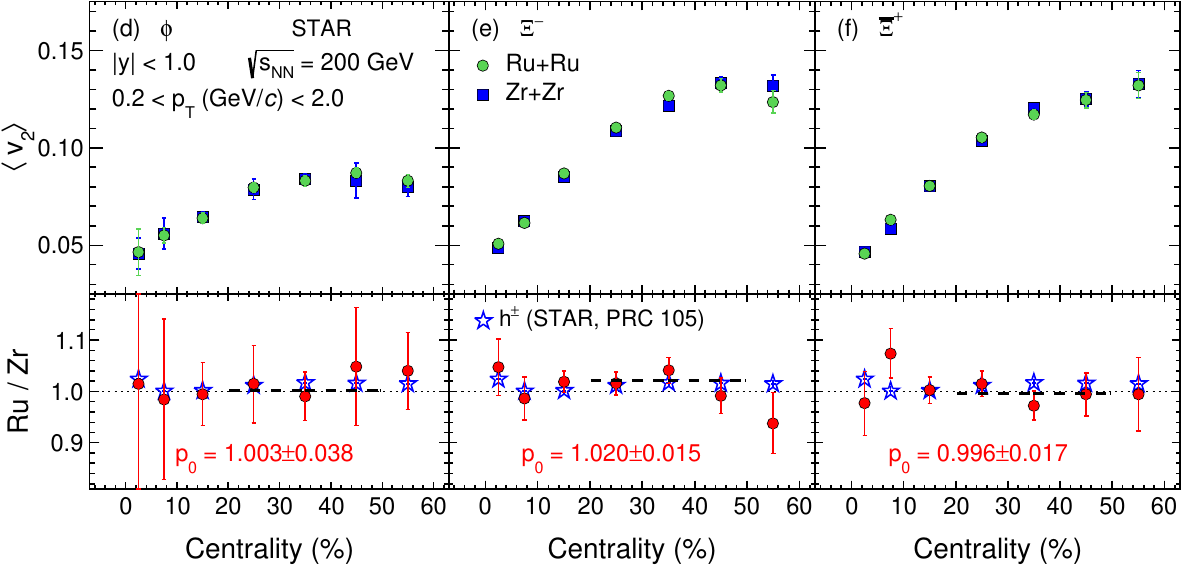}
\caption{\small{$\langle\vtwo\rangle$ as a function of centrality in Ru+Ru and Zr+Zr collisions at $\sqrt{s_{NN}}$ = 200 GeV. The bottom panels show the ratio of $\langle\vtwo\rangle$ between the two isobars. Error bars represent systematic and statistical uncertainties added in quadrature. The dotted lines are the fit to the ratio with a constant polynomial in the centrality range of 20-50\%. The open star markers represent the $\langle\vtwo\rangle$ ratio of charged hadrons~\cite{STARisobarData}.}}
\label{fig:intv2}
\end{figure*}

The structure of the colliding nuclei, including their shapes and sizes, can also affect the elliptic flow. We explore the difference in nuclear structure between the two isobars and its effect on the elliptic flow by studying the centrality dependence of $\pt$-integrated $\vtwo$. Figure~\ref{fig:intv2} shows the $\pt$-integrated $\vtwo$ ($\langle\vtwo\rangle$) as a function of collision centrality in Ru+Ru and Zr+Zr collisions at $\snn$ for strange and multi-strange hadrons. The integration is performed over the $\pt$ range of 0.2-2.0~GeV/$\it{c}$ and $|y| <$ 1.0 to obtain the $\langle\vtwo\rangle$. The $\langle\vtwo\rangle$ is calculated with the yield of the particles not corrected for detector effects. The $\langle\vtwo\rangle$ increases from central to peripheral collisions as inferred from the centrality dependence of $\vtwo$($\pt$). To probe the effect of nuclear structure on $\vtwo$, we analyze the ratio of $\langle\vtwo\rangle$ between the two isobars. The bottom panels in Fig.~\ref{fig:intv2} show the $\langle\vtwo\rangle$ ratio between Ru+Ru and Zr+Zr collisions. The strange hadrons, $K_{s}^{0}$, $\Lambda$, and $\overline{\Lambda}$ show a clear deviation from unity by $\sim$2\% in most-central and mid-central collisions. The deviation in most central collisions can be a result of the larger quadrupole deformity in Ru ($\beta_{2}$ = 0.162) compared to the Zr ($\beta_{2}$ = 0.06) nucleus~\cite{myampt}. To further study the deviation observed in mid-central collisions, the ratio between Ru+Ru and Zr+Zr collisions is fitted with a constant polynomial in the 20-50\% centrality range. The fit values are 1.026$\pm$0.007 for $K_{s}^{0}$ and 1.028$\pm$0.003 (1.020$\pm$0.003) for $\Lambda$ ($\overline{\Lambda}$). This indicates a difference in the nuclear density between the isobar pair, resulting in a larger eccentricity in mid-central Ru+Ru collisions than in Zr+Zr collisions~\cite{STARisobarData,DFT-1,DFT-2}.  These $\langle\vtwo\rangle$ ratios are significantly above unity by 3.7$\sigma$ for $K_{s}^{0}$ and 9.3(6.7)$\sigma$ for $\Lambda$ ($\overline{\Lambda}$). These ratios have also been compared with the $\langle\vtwo\rangle$ ratio of charged hadrons in isobar collisions at $\snn$ which shows a similar trend of $\langle\vtwo\rangle$ with centrality, and the magnitude of deviation is consistent with that of charged hadrons within uncertainties~\cite{STARisobarData}. For the multi-strange hadrons $\phi$ and $\Xi$, the ratio is unity within the uncertainties for the current statistics. The results for $\Omega$ baryons are not presented due to the larger uncertainties.

\subsection{System size dependence}
A wealth of data is available from various collision systems such as U+U, Au+Au, Ru+Ru, Zr+Zr, and Cu+Cu at top RHIC energy~\cite{flowe3,AuCent,STARUU}. These colliding species vary in mass number, nuclear size and the corresponding number of participants ($N_\mathrm{part}$). The increase in nuclear size at a given centrality would increase the $N_\mathrm{part}$ within the initial overlap region, which may result in a larger elliptic flow of particles. To investigate the effect of these nuclear dynamics, we compare $\vtwo$ of strange and multi-strange hadrons in isobar collisions with other collision systems at similar collision energy. 

\begin{figure*}[!htbp]
\centering
\includegraphics[totalheight=7.9cm]{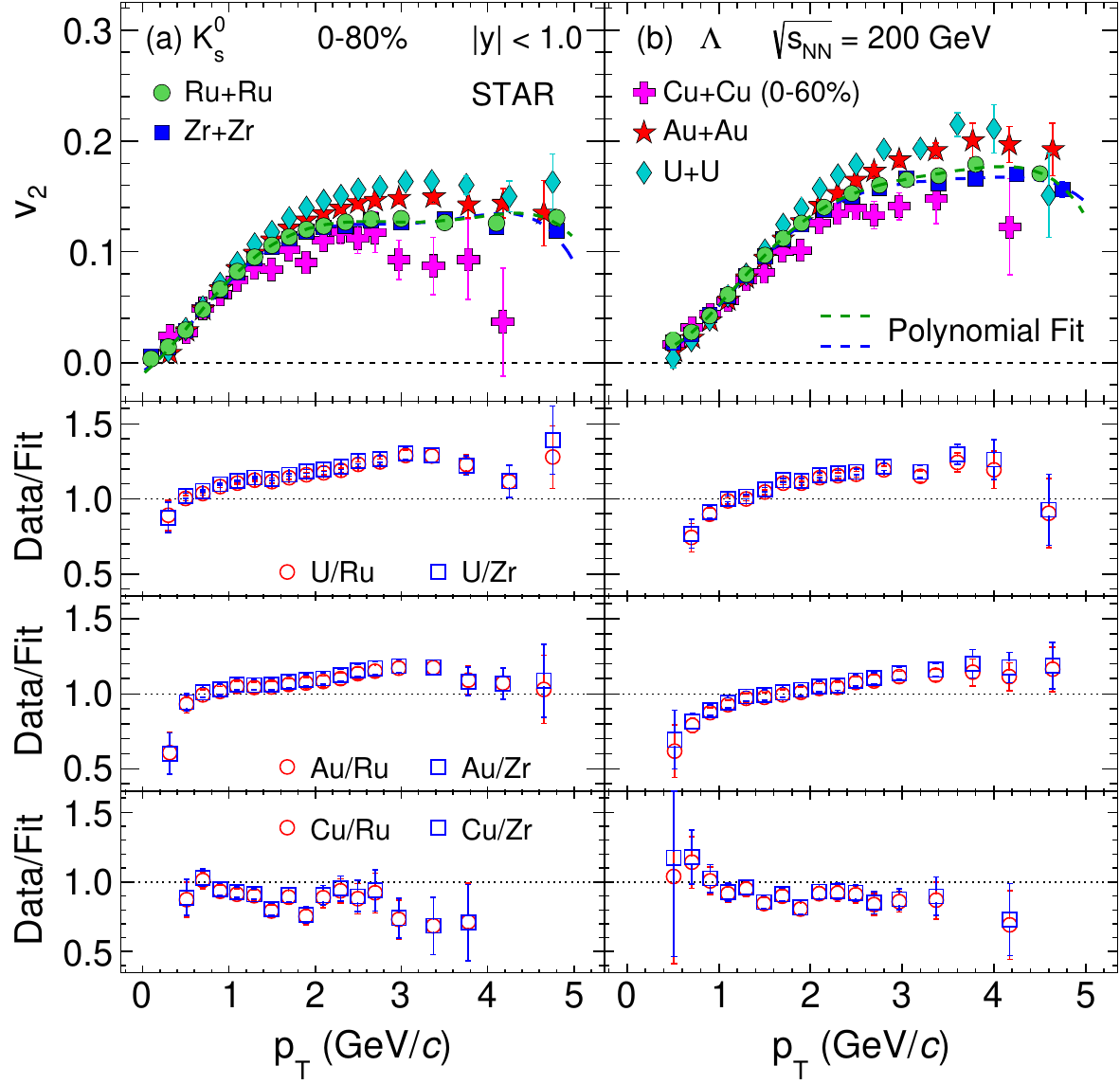}
\includegraphics[totalheight=7.9cm]{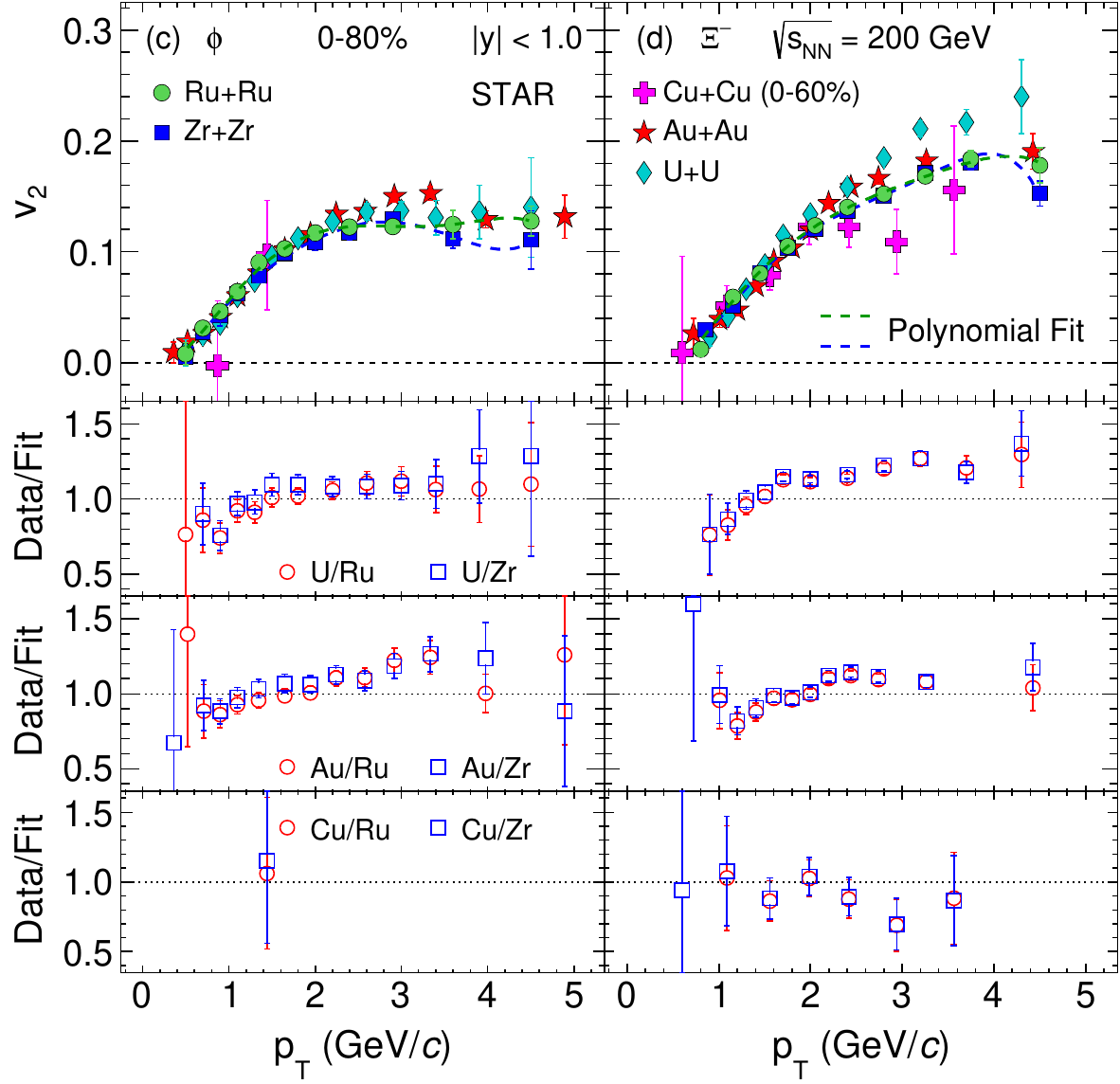}
\caption{\small{$\vtwo$ as a function of $\pt$ in Cu+Cu, Ru+Ru, Zr+Zr, and Au+Au collisions at $\sqrt{s_{NN}}$ = 200 GeV and U+U collisions at $\sqrt{s_{NN}}$ = 193 GeV. The $\vtwo$ from isobar collisions are fitted with a fifth-order polynomial function represented by dashed lines. The bottom panels show the $\vtwo$ ratio between a given system and isobars. Error bars represent statistical and systematic uncertainties added in quadrature.}}
\label{fig:syssize}
\end{figure*}

Figure~\ref{fig:syssize} shows a comparison of $\vtwo$($\pt$) at mid-rapidity for $K_{s}^{0}$, $\Lambda$, $\phi$, and $\Xi$ in minimum bias Cu+Cu, Au+Au, Ru+Ru, Zr+Zr collisions at $\snn$ and U+U collisions at $\sqrt{s_{\mathrm {NN}}} = 193\mathrm{~GeV}$. The $\vtwo$ in isobar collisions are fitted to a fifth-order polynomial function to extract the ratio of $\vtwo$ with other systems. The bottom panels of Fig.~\ref{fig:syssize} show these $\vtwo$ ratios. We observe an increase in (multi-)strange hadron $\vtwo$ at higher $\pt$ ($>$ 1.5 GeV/c) following the hierarchy, $\vtwo^{\mathrm{Cu}} < \vtwo^{\mathrm{Ru}}\approx\vtwo^{\mathrm{Zr}} < \vtwo^{\mathrm{Au}} < \vtwo^{\mathrm{U}}$ with increasing system size. This suggests an increase in collectivity with increasing nuclear size. The $\vtwo$ of the $\phi$ meson shows a relatively weaker system size dependence. The ratios show a systematic increase in $\vtwo$ with $\pt$ for larger nuclei.

\subsection{Model comparison}
\label{model}
In order to understand the effect of nuclear structure on the $v_{2}$ of (multi-)strange hadrons in isobar collisions, the results are compared with a multi-phase transport (AMPT) model. The AMPT model version 2.26t9 in string melting (SM) mode with a partonic cross-section of 3 mb is used~\cite{AMPT}. The AMPT model comprises different stages to simulate heavy-ion collisions. First, the initial conditions in AMPT are modeled using the hard minijet partons and soft strings from the HIJING model~\cite{HIJING}. These minijet partons and excited strings fragment into partons in the string melting mode. Further, Zhang's parton cascade (ZPC) model is used to simulate cascading among the partons~\cite{ZPC}. A quark coalescence model is used to combine partons into hadrons in the hadronization process. The interactions among the resulting hadrons are based on a relativistic transport (ART) model~\cite{ART}. Finally, after a hadronic interaction time of 30 fm/$\it{c}$, the phase space information of hadrons are obtained.

The distribution of nucleons inside the nuclei in AMPT is modeled using the Woods-Saxon distribution function defined as follows, 
\begin{equation}
\rho(r,\theta) = \frac{\rho_{0}}{1+e^{{\lbrace[ r - R(\theta)]/a}\rbrace}},
\end{equation}
where $\rho_{0}$ is the normal nuclear density, $r$ is the distance from the centre of the nucleus, $a$ is the surface diffuseness parameter, and $R(\theta,\phi)$ characterizes the deformation of the nucleus, 
\begin{equation} 
R(\theta) = R_{0}[1 + \beta_{2}Y_{2,0}(\theta) + \beta_{3}Y_{3,0}(\theta)].
\end{equation} 
$R_{0}$ represents the effective nuclear radius, $\beta_{2}$ and $\beta_{3}$ are the quadrupole and octupole deformities, and $Y_{l,m}(\theta)$ are the spherical harmonics. The model has been modified to incorporate the deformed Ru and Zr nuclei. Recent AMPT model studies using various WS parameterisations, including a quadrupole and octupole deformation in isobar nuclei, describe the experimental results of the charged hadron $\langle\vtwo\rangle$ ratio at $\snn$~\cite{myampt,PRCallWS}. We study Ru+Ru and Zr+Zr collisions at $\sqrt{s_{\mathrm {NN}}}$ = 200 GeV from the AMPT model with the WS parameterisation as shown in Table~\ref{tab1}~\cite{myampt,PRCallWS}.   

\begin{table}[!htbp]
\begin{center}
\caption{Deformation configuration for the Ru and Zr nuclei in the AMPT model.}
\vspace{0.2cm}
\label{tab1}
\begin{tabular}{ccccc}
\hline
\hline
System & $R_{0}$ & $a$ & $\beta_{2}$ & $\beta_{3}$ \\
\hline
Ru &  5.090 & 0.460 & 0.162 & 0.0 \\
Zr &  5.090 & 0.520 & 0.060 & 0.2 \\
\hline
\hline
\end{tabular}
\end{center}
\end{table}

\begin{figure*}[!htbp]
\centering
\includegraphics[totalheight=7.cm]{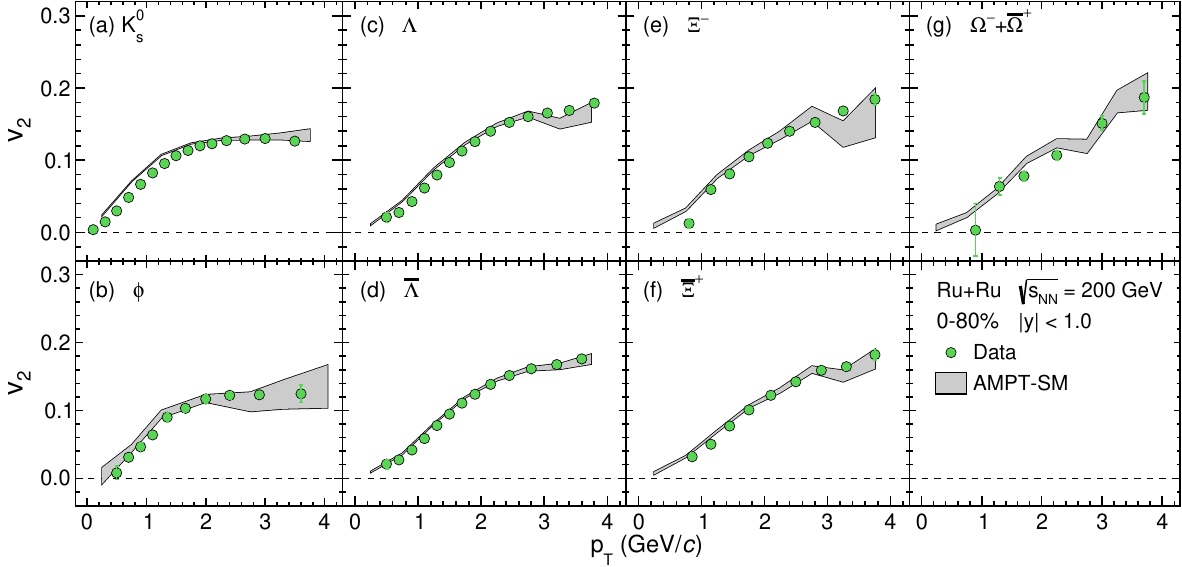}
\caption{\small{Comparison of $\vtwo$ as a function of $\pt$ in minimum-bias Ru+Ru collisions at $\snn$ with the AMPT-SM model.}}
\label{fig:ruampt}
\end{figure*}

\begin{figure*}[!htbp]
\centering
\includegraphics[totalheight=7.cm]{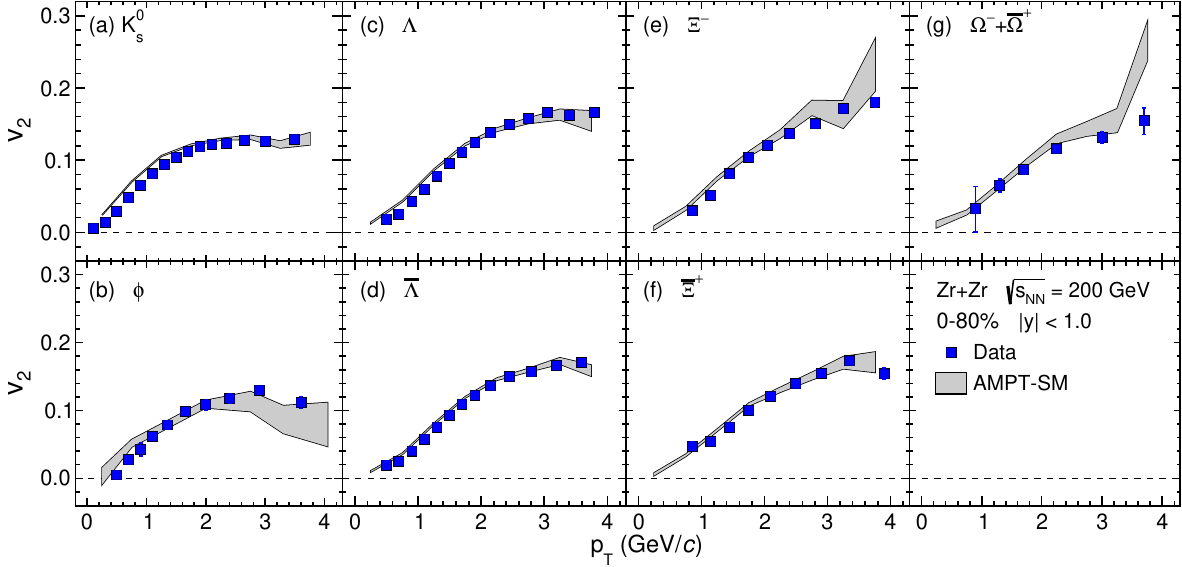}
\caption{\small{Comparison of $\vtwo$ as a function of $\pt$ in minimum-bias Zr+Zr collisions at $\snn$ with the AMPT-SM model.}}
\label{fig:zrampt}
\end{figure*}

A total of 20 million events for Ru+Ru and Zr+Zr collisions at $\sqrt{s_{\mathrm {NN}}}$ = 200 GeV are generated for each of the two collision systems. Figures~\ref{fig:ruampt} and ~\ref{fig:zrampt} show the AMPT-SM model calculations of $\vtwo$($\pt$) for strange and multi-strange hadrons at mid-rapidity compared with the experimental data of minimum-bias Ru+Ru and Zr+Zr collisions at $\snn$, respectively. The model calculations qualitatively describe the experimental data in the measured $\pt$ range in both collision systems.

\section{Summary}
\label{summary}
We report the first differential measurement of strange and multi-strange hadrons elliptic flow at mid-rapidity in large statistics data set of isobar, $^{96}_{44}$Ru+$^{96}_{44}$Ru and $^{96}_{40}$Zr+$^{96}_{40}$Zr, collisions at $\snn$. The transverse momentum dependence of $\vtwo$ for various centrality classes (0-80\%, 0-10\%, 10-40\%, and 40-80\%) is reported. The universal number of constituent quark scaling of $v_{2}$($\pt$) is observed to hold within $\sim$20\% uncertainty for the strange and multi-strange hadrons in Ru+Ru and Zr+Zr collisions, consistent with the observed $n_{q}$ scaling in Au+Au and U+U collisions at top RHIC energy. This suggests the dominance of partonic collectivity and quark coalescence as the particle production mechanism even in these small-sized isobar collisions at intermediate $\pt$. The comparison of $\vtwo$($\pt$) of strange and multi-strange hadrons amongst various collision systems at similar collision energies is also reported. A systematic increase of $\vtwo$($\pt$) in the intermediate $\pt$ indicates higher collectivity with increasing system size. 

Additionally, our findings demonstrate that the $\langle v_{2}\rangle$ ratios of strange hadrons ($K_{s}^{0}$, $\Lambda$ and $\overline{\Lambda}$) between the isobar systems differ from unity by $\sim$2\% in the central and mid-central collisions. This difference in the mid-central collisions can be attributed to the surface diffuseness. The difference in the ratio in central collisions indicates at the higher quadrupole deformation in Ru nuclei than the Zr nuclei. The ratio for $\phi$ and $\Xi$ are unity within the current uncertainties. The AMPT-SM model qualitatively describes the $\pt$-differential $\vtwo$ measurements for all the studied particles in Ru+Ru and Zr+Zr collisions at $\snn$. 

\section{ACKNOWLEDGEMENTS}
We thank the RHIC Operations Group and SDCC at BNL, the NERSC Center at LBNL, and the Open Science Grid consortium for providing resources and support.  This work was supported in part by the Office of Nuclear Physics within the U.S. DOE Office of Science, the U.S. National Science Foundation, National Natural Science Foundation of China, Chinese Academy of Science, the Ministry of Science and Technology of China and the Chinese Ministry of Education, NSTC Taipei, the National Research Foundation of Korea, Czech Science Foundation and Ministry of Education, Youth and Sports of the Czech Republic, Hungarian National Research, Development and Innovation Office, New National Excellency Programme of the Hungarian Ministry of Human Capacities, Department of Atomic Energy and Department of Science and Technology of the Government of India, the National Science Centre and WUT ID-UB of Poland, the Ministry of Science, Education and Sports of the Republic of Croatia, German Bundesministerium f\"ur Bildung, Wissenschaft, Forschung and Technologie (BMBF), Helmholtz Association, Ministry of Education, Culture, Sports, Science, and Technology (MEXT), and Japan Society for the Promotion of Science (JSPS), and Agencia Nacional de Investigaci\'on y Desarrollo de Chile (ANID), Chile.

\end{document}